\documentclass[useAMS,usenatbib]{mn2e}
\usepackage{graphicx}
 \usepackage{times}
 \usepackage{rotating}
\usepackage{txfonts}
 \usepackage{rotating}
 \usepackage{lscape}
\usepackage{booktabs}
\usepackage{longtable}
\usepackage{enumerate}
\usepackage[T1]{fontenc}
\usepackage{aecompl}

\def\vb{\hbox{OGLE004633.76$-$731204.3}}

\newcommand{\apj}{ApJ}
\newcommand{\apjs}{ApJS}

\newcommand{\aap}{A{\&}A}
\newcommand{\ARep}{ARep}
\newcommand{\aaps}{A{\&}AS}
\newcommand{\mnras}{MNRAS}
\newcommand{\aj}{AJ}
\newcommand{\araa}{ARAA}
\newcommand{\pasp}{PASP}
\newcommand{\na}{New Astronomy}
\newcommand{\nat}{Nature}

\newcommand{\actaa}{AcA}

\DeclareGraphicsExtensions{.gif,.pdf,.bmp,.jpg}

 \def\gtrsim{\mathrel{\hbox{\rlap{\hbox{\lower4pt\hbox{$\sim$}}}\hbox{$>$}}}}
 \def\ltsim{\mathrel{\hbox{\rlap{\hbox{\lower4pt\hbox{$\sim$}}}\hbox{$<$}}}}

\title[The eclipsing LMC binary ELHC\,10]{ On the eclipsing binary ELHC~10 with occulting dark disc in the Large Magellanic Cloud}

\author[H.E. Garrido, R.E. Mennickent, G. Djura{\v s}evi{\'c} et al.]
  {H. E. Garrido,$^{1}$\thanks{E-mail: hgarrido@ucsc.cl}
  R. E. Mennickent$^{2}$, G. Djura{\v s}evi{\'c}$^{3,4}$,   L. Schmitdtobreick$^{5}$, D. Graczyk$^{2,6}$, \newauthor S. Villanova$^{2}$, D. Barr\'ia$^{2}$ \\
  $^{1}$ Universidad Cat\'olica de la Sant\'isima Concepci\'on, Departamento de Matem\'atica y F\'isica Aplicadas, Concepci\'on, Chile.\\
 $^{2}$ Universidad de Concepci\'on, Departamento de Astronom\'{\i}a, Casilla 160-C, Concepci\'on, Chile.\\
  $^{3}$ Astronomical Observatory, Volgina 7, 11060 Belgrade 38, Serbia.\\
  $^{4}$ Issac Newton Institute of Chile, Yugoslavia Branch, Belgrade, Serbia.\\
   $^{5}$European Organisation for Astronomical Research in the Southern Hemisphere, Alonso de Cordoba 3107, Vitacura, Casilla 19001, Santiago 19, Chile.\\
   $^{6}$Millenium Institute of Astrophysics, Av. Vicu\~{n}a Mackenna 4860, Santiago, Chile\\ }
\date{}

\pagerange{\pageref{firstpage}--\pageref{lastpage}} \pubyear{2012}

\def\LaTeX{L\kern-.36em\raise.3ex\hbox{a}\kern-.15em
    T\kern-.1667em\lower.7ex\hbox{E}\kern-.125emX}

\begin{document}

\label{firstpage}

\maketitle

\begin{abstract}
{ We investigate the luminous star ELHC 10 located in the bar of the Large Magellanic Cloud, concluding that it is a SB1 long-period eclipsing binary where the main eclipse is produced by an opaque structure hiding the secondary star. For the more luminous component  we determine an effective temperature of 6500 $\pm$ 250 $K$, log\,g  = 1.0 $\pm$ 0.5 and luminosity 5970 L$_{\sun}$. From the radial velocities of their photospheric lines we
calculate a mass function of  7.37 $\pm$ 0.55 M$_{\sun}$. Besides Balmer and forbidden N {\sc ii} emission, we find splitting of metallic lines, characterized by strong discrete absorption components (DACs), alternatively seen at the blue and red side of the photospheric spectrum. These observations hardly can be interpreted in terms of an structured atmosphere but might reflect mass streams in an interacting binary. The primary shows signatures of s-process nucleosynthesis and might be a low-mass post-AGB star with a rare evolutionary past if the binary is semi-detached.  The peak separation and constancy of radial velocity in H$\alpha$ suggest that most of the Balmer emission comes from a circumbinary disc. }   
 \end{abstract}


\section{Introduction}

ELHC~10 (also known as [BE74] 561, 
2MASS J05194770-6939121 or OGLE LMC-LPV-41682)  is a binary system with period of 219.9 d\footnote{http://ogledb.astrouw.edu.pl/$\sim$ogle/CVS/}, located near the nebular complex N\,120  \citep{Henize1956} in the bar of the Large 
Magellanic Cloud (LMC); it was identified  as a H$\alpha$ emission-line star by \citet{Bohannan1974}.  The name ELHC\,10 for this object was proposed by \citet{Lamers1999} and 
\citet{deWit2002}, who searched the EROS database for blue objects with irregular photometric behavior similar to Galactic Herbig Ae/\,Be stars in the LMC bar. ELHC~10 was the 
reddest and brightest star in their sample, with a color $(B -V ) = 0.30$ and  estimated absolute magnitude of $M_{v} = -4.7$. The low resolution spectrum showed emission 
in H${\alpha}$ with equivalent width of -23 \AA. Contrary to the other stars in their sample, this star was not an early B star;
the spectrum was more compatible with an early F-type giant or supergiant. The spectral type  was estimated  as F2/5 I-II, and the object was included in a sample of HAeBe stars \citep{deWit2005}. Although Herbig Ae/ Be stars (or HAeBe) are emission-line stars of spectral types O, B, A and in a few cases F, 
and in most instances exhibit IR excesses, which are attributed to dust emission from circumstellar discs, this star is hardly a HAeBe candidate, due to the much 
evolved nature of the luminous component.

In this paper, we  focus on resolving the uncertainty about the nature of this system by performing an analysis of the stellar components and the surrounding environment. High-resolution spectra and very accurate $BRI$  light curves allowed us to study  ELHC 10  in unprecedented detail for the first time.  We  find that the brightest member of this binary has a low surface gravity and an atmosphere depleted in refractory elements and might be identified as a member of the subclass of post-Asymptotic Giant Branch stars. In addition, we provide evidence for the existence of a circumbinary disc and a circumstellar opaque disc.

There are many substantial evidences for stable and compact Keplerian circumbinary disks around many post-AGB objects \citep{ de Ruyter2006, Gielen2008, Gielen2009, vanAarle2011, Hillen2013, Hillen2014, Hillen2015, Bujarrabal2015}. Additionally the binary nature of the disc source has been in some cases  confirmed, with the most famous example certainly being HD 44179, better known as the Red Rectangle Nebula (\citealt{de Ruyter2005, Witt2009, Thomas2012, Martínez González 2015}; \citealt[][and references therein]{van Winckel 1995}).  Two other important cases  where it has been possible to test the existence of stable Keplerian discs and the correlation with binarity in post-AGB stars have been reported by \cite{Gorlova2012, Gorlova2015}. Hence, the present work aims to inquire into the nature of ELHC 10 bringing new light on the formation of stable  discs in these binary systems, which are produced presumably  by means of interaction with the companion, when the more massive component becomes either a red giant (RG) or an AGB star. Actually, these amazing and rare systems might represent a new evolutionary channel  that until now has been poorly studied.
 
In  Section 2 we present a detailed report of the photometric and spectroscopic observations used in this paper. In Section 3 we analyze these observations to obtain the orbital ephemeris and a preliminary spectral type for the more luminous  component.
The spectral energy distribution is also analyzed in this section and a radial velocity study is presented. A study of line profiles yields 
to detailed parameters  of the more prominent star. In  Section 4 we use results of our radial velocity study to get information about masses and offer an interpretation of discrete absorption components (DACs) in terms of mass exchange and mass loss in an interacting binary. Finally, in Section 5 we provide the conclusions of our work. 

In this paper we call the primary star the stellar component whose absorption lines are detected in the spectrum,  while the secondary  is the undetected (and more massive) star. Let's remember that this scenario of a less evolved more massive star can be possible in case of past or present mass transfer by Roche-lobe overflow, as happens in Algols. Quantities relative to these stars are labeled with subindexes {\it 1} and {\it 2}, respectively.

\begin{figure*}
    \centering
    \includegraphics[width=0.8\textwidth]{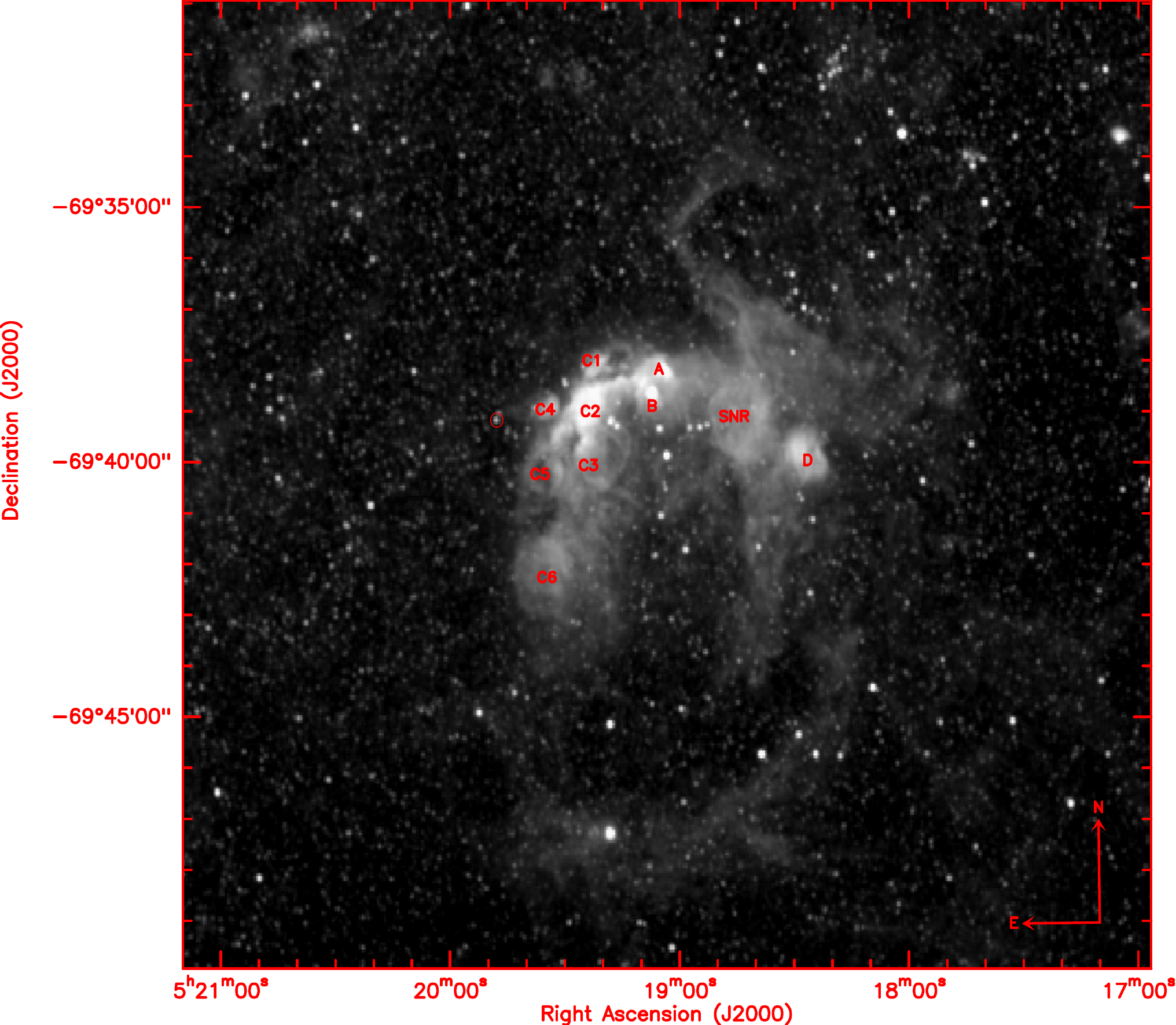}
    \caption[H$\alpha$ image from MCELS on the center of nebular complex N120.]{H$\alpha$ image from MCELS on the center of nebular complex N120. The position  of ELHC\,10  is marked with red circle  in the absolute coordinates $\alpha(2000)$= 05:19:47.70 and $\delta (2000)=-$69:39:12.1. 
    See text for more details.}
    \label{fig1}
\end{figure*}

\begin{table*}\label{Tabla0}
\caption{Summary of new spectroscopic observations. The HJDs at mid-exposure are given. Phases refer to the ephemeris given in Eq.\,1. The signal to noise ratio is calculated at the continuum around H$\alpha$.
We report the visibility of discrete absorption component in the NaD line as discussed in the text.}
\begin{tabular}{llcccccccc} 
\hline
Phase & Telescope & Instrument & Date-obs & UT-start & exptime (s) & Airmass & HJD &  S/N  & DACs \\
\hline
0.04& ESO/3.6 & HARPS & 2014/11/06  &04:19:02 & 2400 & 1.419 & 2456967.693794 &   14 & BACs \\
0.22& ESO/3.6 & HARPS & 2010/09/28  &05:05:09 & 2500 & 1.737 & 2455467.712230 &  20 & RACs \\
0.25 & ESO/3.6 & HARPS & 2013/10/08  &05:44:38 & 2000 & 1.505 & 2456573.739580 &  22& RACs\\
0.47 & LCO/Clay& MIKE & 2012/02/07 &00:22:03 & 900 &1.323 & 2455964.514824 & 25 &No  \\
0.54 & LCO/Clay& MIKE & 2013/12/11 & 00:21:28& 600 &1.718 & 2456637.518113 & 25&No  \\
0.54 & LCO/Clay& MIKE & 2013/12/12 & 00:17:46 & 600 &1.717 & 2456638.515537 & 25&No   \\
0.62 & Ir\'en\'ee Du Pont & ECHELLE & 2010/12/27 & 05:36:25 & 900 & 1.387 & 2455557.743669&  40&BACs\\\hline
\end{tabular}
\vspace{0.2 cm}
\end{table*}

\begin{table*}\label{Tabla1}
\caption{EROS-2 and OGLE II photometry, dereddened B-V color and derived spectral type for ELHC\,10. }
\begin{tabular}{ccccccccc}
\hline
B$_{\mathrm E}$&R$_{\mathrm E}$&B& V  & I & (B-V) & (V-I) & (B-V)$_{0}$ & Spectral Type \\\hline
13.576 &13.503 &14.428& 13.931 & 13.484 & 0.494 & 0.446 & 0.32 &  F5 \\\hline
\end{tabular}
\end{table*}


\section{Observations and data reduction}

\subsection{Spectroscopy data}

We conducted optical spectroscopic observations of ELHC\,10 at different epochs during 4 years with several echelle spectrographs. Three  high-resolution spectra were obtained in September 2010, October 2013 and November 2014,
using the HARPS Echelle Spectrograph mounted on the ESO 3.6 m telescope at La Silla Observatory, Chile. The  two CCD chips on this instrument  sample the spectral range of  3780-6910 \AA.  The fibres  
allowed us to obtain spectra at resolving power $\approx 115 000$. The HARPS spectra do not include the 5304 to 5337 $\AA$ region, because of the gap between the two detectors. 
 
Three additional spectra were secured with the Magellan Inamori Kyocera Echelle (MIKE) spectrograph at the Clay Telescope in Las Campanas Observatory, Chile, in February 2012  and  December 2013. This double echelle
spectrograph provided wavelength coverage of 3390-4965 \AA\  (blue camera) and 4974-9407 \AA \ (red camera). With a slit width of 0.7 arcsec, the resolving power was 40000.  Unfortunately, the two spectra obtained in
December 2013 could not be calibrated in wavelength, but they  were used to perform a qualitative analysis of line shapes.

 We also obtained one high-resolution ($R\approx$ 40000) spectrum with the DuPont Echelle Spectrograph at Las Campanas Observatory, covering the spectral region 4448-7982 \AA\  in December of 2010. 
 
 All spectra discussed in this work are corrected for earth translational motion and normalized to the continuum. The radial velocities (RVs) are heliocentric.  The data reductions were  done with standard 
 {\it IRAF}\,\footnote{IRAF is distributed by the National Optical Astronomy Observatories, which are operated by the Association of Universities for Research in Astronomy, Inc., under cooperative agreement with the National
 Science Foundation.} routines  for echelle spectroscopy, including flat and bias correction, wavelength calibration and order merging. No flux calibration was needed for our purposes. A summary of our spectroscopic
 observations is shown in Table \ref{Tabla0}.
 
 \subsection{Photometric data}

We used the H$\alpha$ images from the Magellanic Cloud Emission-Line Survey \citep{Smith1999} to get information about the conditions surrounding the environment of ELHC\,10. As mentioned above, ELHC\,10 is located
near the nebular complex N\,120. This complex is composed of smaller nebulae arranged in a bright incomplete ring with the exciting stars  in the centre of the bubble-like nebular complex. Firstly, Henize in 1956 
separated the nebula into four brighter overlapping H\,{\sc ii} regions: N\,120 A, B, C and D. Additional  images of this region as those of MCELS, revealed that N\,120 C was composed of smaller nebulae, denoted C1 to C6,
as shown in Fig. \ref{fig1}. The heliocentric velocity of the nebulae C1 to C6 {\bf were accurately  measured to be between 250} and 263 km\,s$^{-1}$, with velocity dispersions between 3 to 7 km\,s$^{-1}$ by \citet{Laval1992}. 
In particular, we are interested in the heliocentric velocities of the nebular complex N\,120 C4 and C5  that are closer to our star; these nebulae have heliocentric velocities between 260 and 261 km\,s$^{-1}$.

Multi-band  and time-series photometry was retrieved from the Optical Gravitational Lensing Experiment in its third phase (OGLE III; \citealt{Udalski2002}), the Exp\'erience de Recherche d'Objects Sombres in its second  phase
EROS-2\footnote{http://eros.in2p3.fr/} and the Massive Compact Halo Object (MACHOs\footnote{http://wwwmacho.anu.edu.au/}) project.

The OGLE III photometry was accumulated  over an interval of almost 8 years, from July 2001 to May 2009, with the 1.3 meter Warsaw telescope located at Las Campanas Observatory in Chile. The majority of these observations
were taken in the $I$-band, typically about 500 points,  but  ELHC\,10 has OGLE II observations too, which increases the number of points up to 1000. About 40-60 observations ($\approx 100$ with the OGLE II data) were secured
in the $V$-band. See \cite{Soszynski2009} for more information about the {\bf obtention of OGLE photometric data.}

The EROS-2 observations  were obtained using the  1 m Ritchey-Chr\'etien Telescope at ESO, La Silla-Chile. The telescope was equipped with two cameras, each camera contained eight $2048\,\times\,2048$ detectors mounted in a mosaic
pattern. The pixel size is $0.6''$, with a typical seeing of $2''$ FWHM at the site. The EROS-2 Photometric System consisted of a  $B_{\mathrm E}$ (420-720 nm, blue) band and $R_{\mathrm E}$ (620-920 nm, red) band. 
The  $B_{\mathrm E}$  band  is close to the Johnson $V$-band, but it is broader. The  $R_{\mathrm E}$ band is intermediate between the bands of Cousins $R$ and $I$. The photometric points obtained using  the $B$-band 
generally have  better accuracy  than in the $R$-band \citep{Grison1995, Ansari2001}.

\begin{figure*}
    \includegraphics[width=0.45\textwidth]{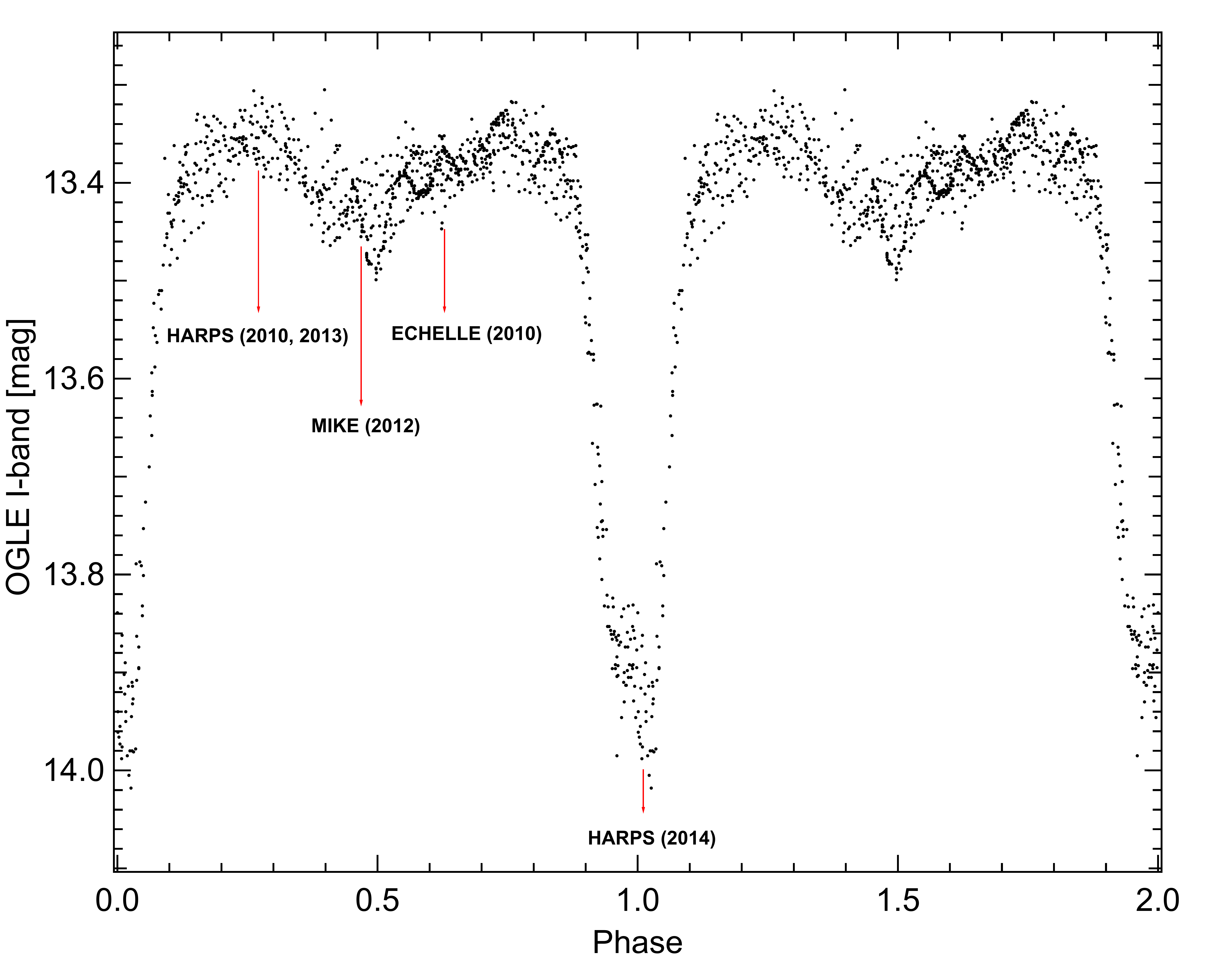}  
        \includegraphics[width=0.45\textwidth]{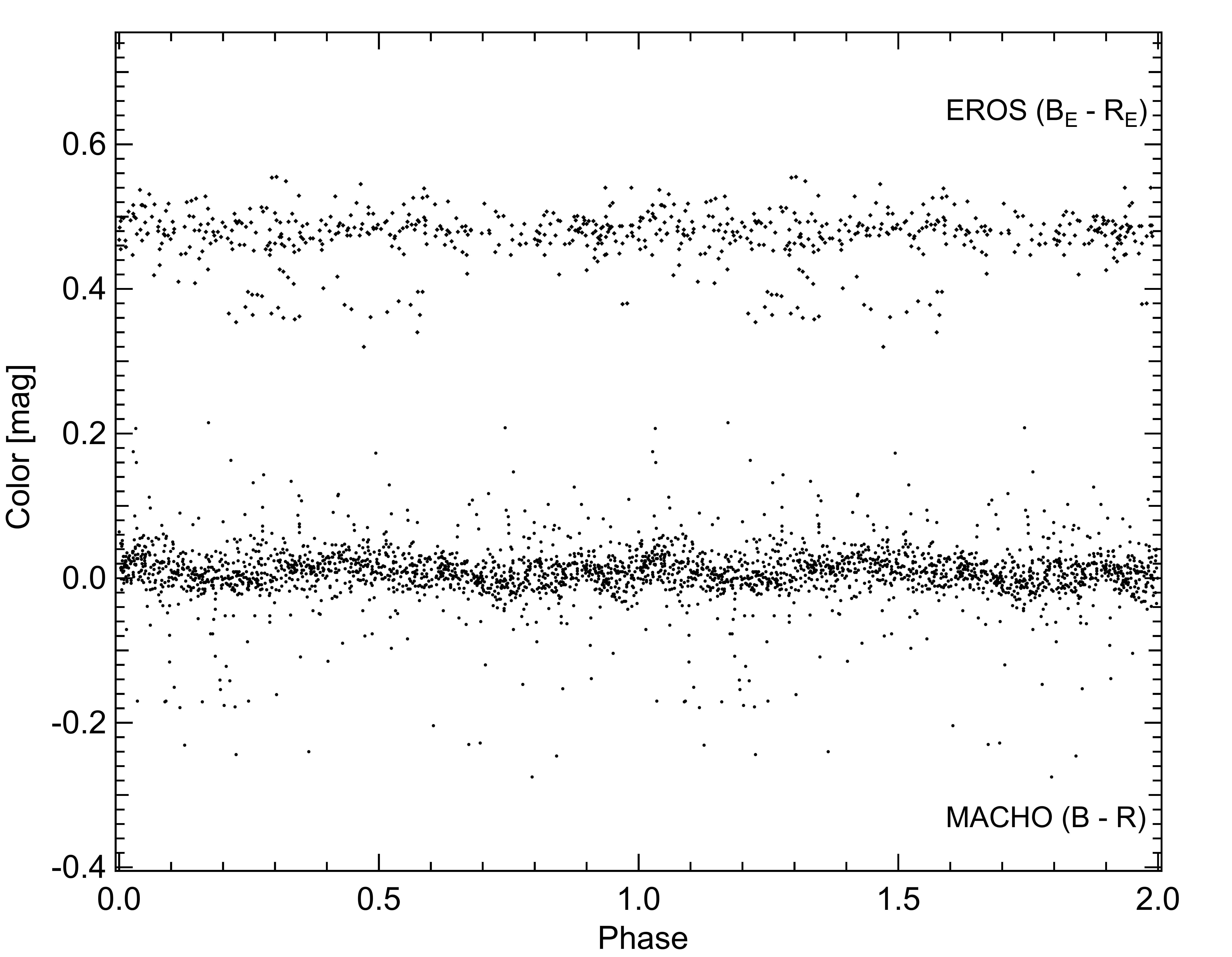}  
   \includegraphics[width=0.45\textwidth]{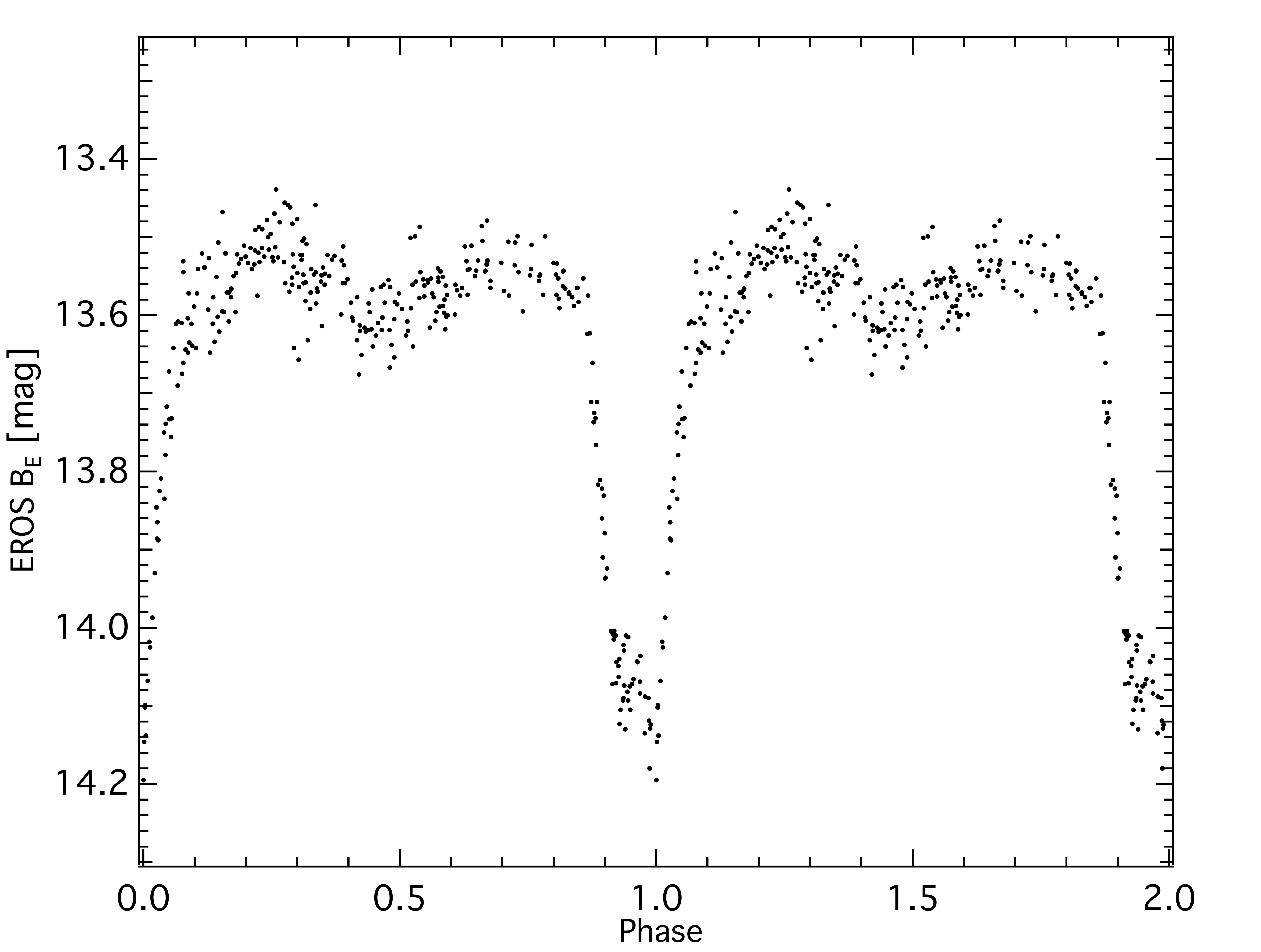} 
      \includegraphics[width=0.45\textwidth]{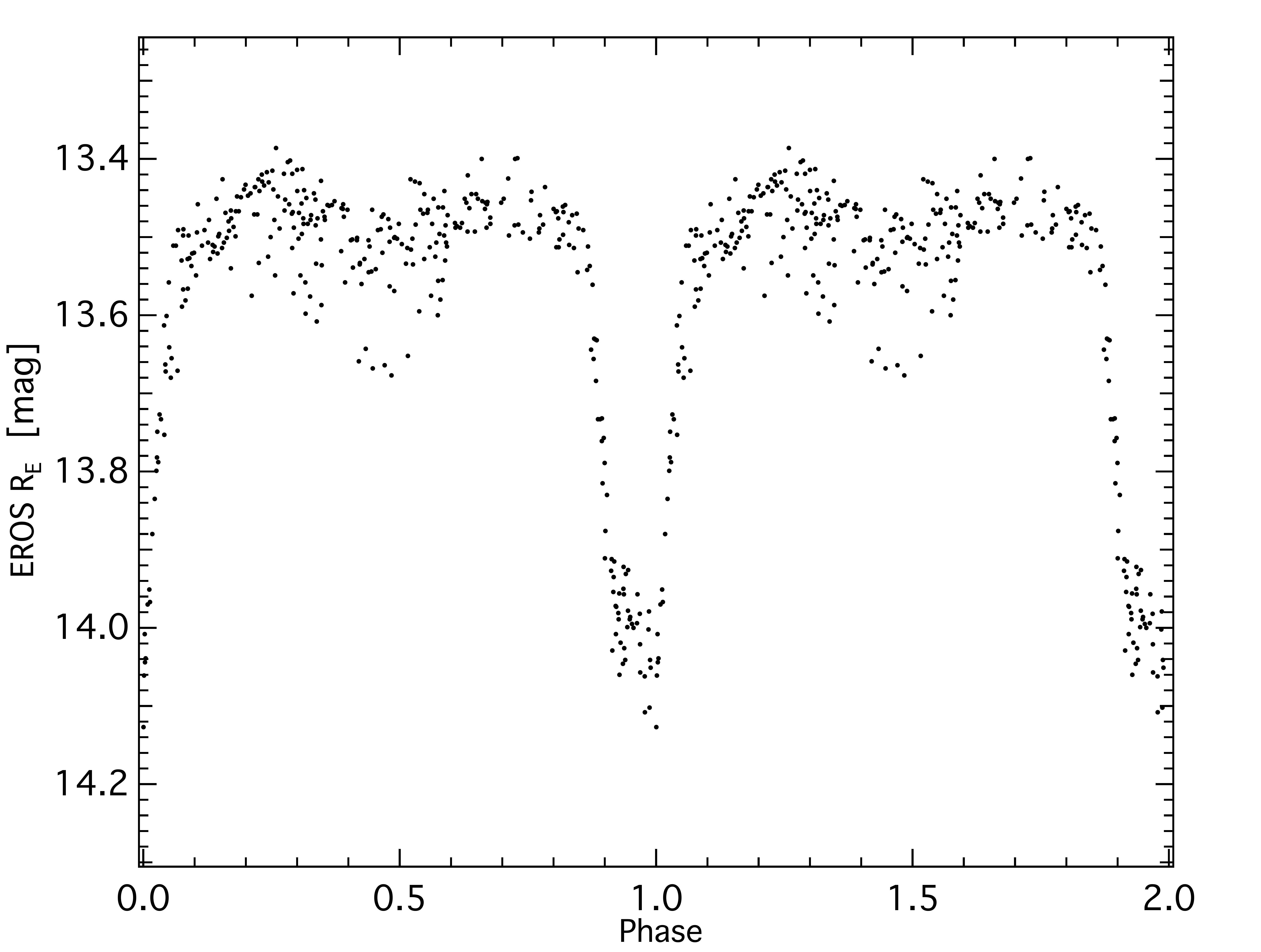} 
    \caption{$I$-band, $B_{\mathrm E}$-band and $R_{\mathrm E}$-band phase diagrams and EROS-2  and MACHO colors for ELHC\,10, showing primary and secondary eclipses. For clarity the phase has been extended over two periods.
    Epochs for our observations are indicated by arrows. EROS-2 colors are shifted by $+$0.4 mag. }
    \label{fig2}
\end{figure*}

 The MACHO survey provides instrumental magnitudes for each star in two contiguous ``blue'' and ``red'' passbands, labeled $B$ and $R$, at different effective wavelengths than the standard B and R passbands in the Johnson-Cousins
 photometric system \citep{Alcock1999}. The MACHO light curves have between 500 to 1300 points per star, depending on the filter  with which they were taken. The MACHO photometry was accumulated over the interval of 8 years
 from July 1992 to January 2000. The light curves retrieved from MACHO together with the EROS-2 light curves were used to construct color curves for ELHC\,10 (Fig. \ref{fig2}).

\begin{figure}
    \includegraphics[width=0.45\textwidth]{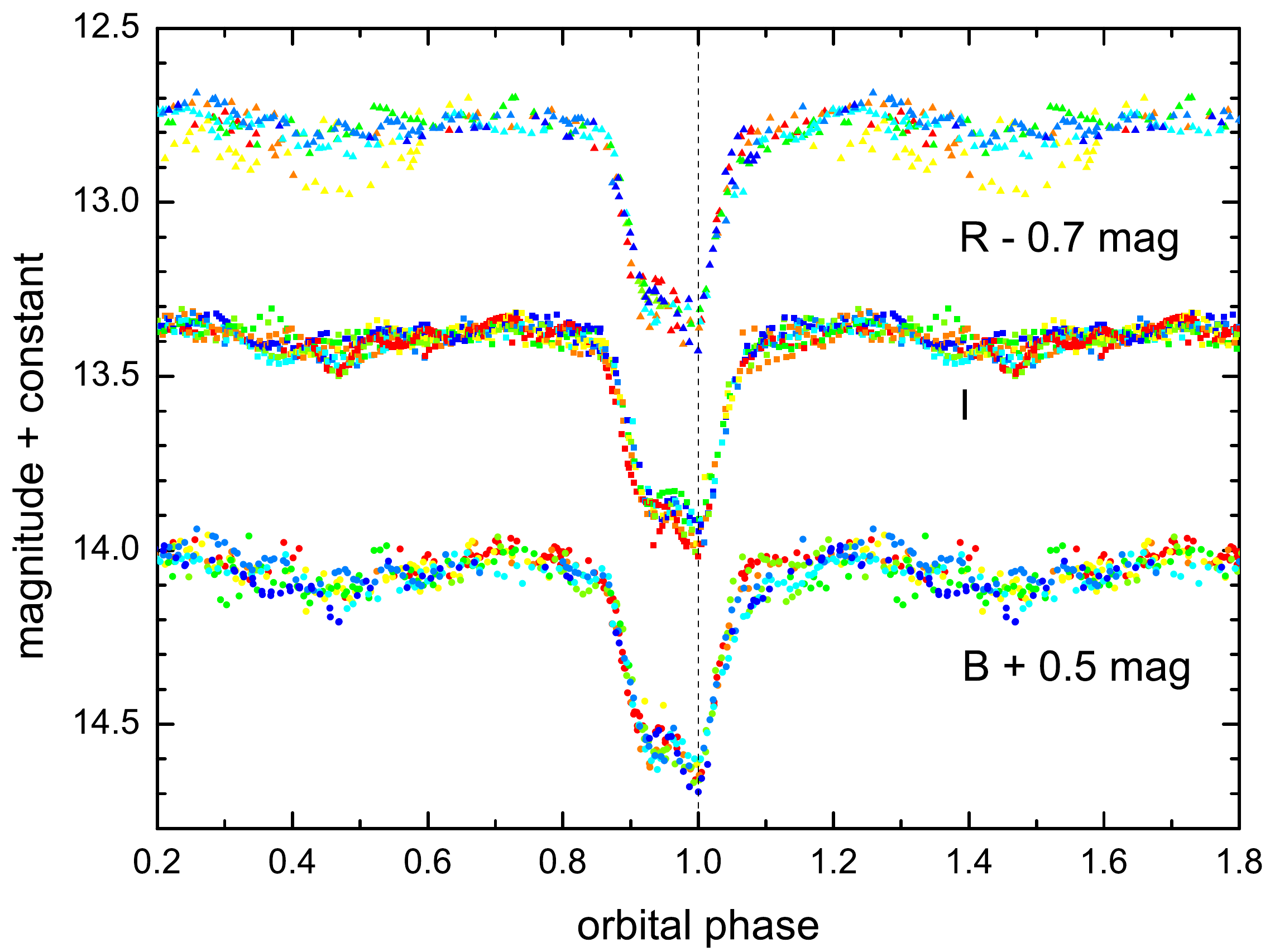}  
            \caption{The main eclipse at different bandpasses. Data points are colored in a sequence changing with time, in order to see sub-orbital variability.}
    \label{varibilityLCs}
\end{figure} 

\section{Results}

\subsection{Photometric characterization and period search}\label{period}

In Table\,2 we give the EROS-2 magnitudes and mean $BVI$ magnitudes and colors obtained from the OGLE II catalog.  OGLE II has a median seeing for the entire dataset of about $1.3''$, and the uncertainty of the zero-point is 
less than 0.02 mag \citep{Udalski2000}. We calculated derredened $(B-V)_{0}$ colors assuming mean internal reddening $E(B-V)= 0.16$ mag given by \cite{Oestreicher1996}, as representative for the gas and dust distribution 
inside the LMC.
 
 The $(B-V)_{0}= 0.32$ color is concordant with  a middle F-type giant or supergiant, and this spectral type determination  is compatible with the lines observed in the HARPS spectra discussed in Section\,3.2 and the earlier 
 estimate by \cite{deWit2005}.

The OGLE III $I$-band and $V$-band light curves have been analyzed utilizing the Phase Dispersion Minimization (PDM) algorithm introduced by \cite{Stellingwerf1978}, as implemented in {\it IRAF}. We have found the
fundamental period of $219.9 \pm 2.4$ days on the $I$-band time series. The period error is the half-width at half-maximum (HWHM) of the periodogram's peak. The ephemeris for the main eclipses is:

\begin{equation}\label{1}
T_{min}(\mathrm{HJD})=2450582.513 + (219.9\,^{d} \pm 2.4)\times E.
\end{equation}

 We have modeled the OGLE III and EROS-2 light curves with a Fourier series including the orbital frequency plus harmonics following the method described by \cite{2012MNRAS.421..862M}. The analysis of the residuals of the light
curve regarding this model did not reveal any additional periodicity.

Afterwards, OGLE III and EROS-2 light curves were folded using the ephemeris given by Eq.\,1, and the result is shown in Fig. \ref{fig2}. The structure seen in the eclipses of ELHC\,10,
in particular the plateau,
is similar to that exhibited in EE\,Cep \citep{Galan2012} and the newly discovered systems OGLE LMC-ECL-11893 \citep{2014ApJ...797....6S, Dong2014}, 
OGLE-BLG182.1.162852 \citep{Rattenbury2015} and  \vb\ \citep{Mennickent2010a}. 
These systems have been described as long-period eclipsing binaries with circumstellar discs and the most plausible hypothesis to explain the observed shape of the light curve is a nearly edge-on dusty disk around a low-luminosity central object, this specific configuration is very similar to the geometry of the eclipses in the $\varepsilon$ Aur system \citep{Kloppenborg2015} and the plateau reflects the fraction covered for the opaque circumsecondary disc onto the primary stellar disc.

The OGLE $V$-band and $I$-band light curves were phased with the 219.9 day period, 
and then interpolated to a constant phase-step to get a representation of the average light variability during the cycle at each band. 
The color $V-I$ was constructed by subtracting both averaged light curves; it does not show
any significant change during the cycle, but remains constant at the value $V-I$ = 0.5, representative of a
F2-F5 type supergiant. The same color constancy  is observed in MACHO and EROS colors (Fig. \ref{fig2}).
Actually, the MACHO color shows the system slightly bluer by 0.01 mag. at quadratures.
The almost gray and irregular eclipse is compatible with occultation of the primary by a  inhomogeneous disk dominated by particles of quite large diameters, mixture of grains and dust similar to the discs in $\varepsilon$ Aur and EE Cep system \citep{Hoard2010, Galan2012}

The variability of the eclipse shape at different epochs and also short-term quasi-periodic oscillations can be observed in Fig.\, \ref{varibilityLCs}.
The scatter observed in the $R$-band at certain phases occurs only during one epoch (yellow points) and near secondary eclipse; it could be instrumental error related to problems with removal of H$\alpha$ luminosity of the background nebulosity or alternatively, reveals changes in H$\alpha$ emission strength.


\begin{figure*}
     \includegraphics[width=0.495\textwidth]{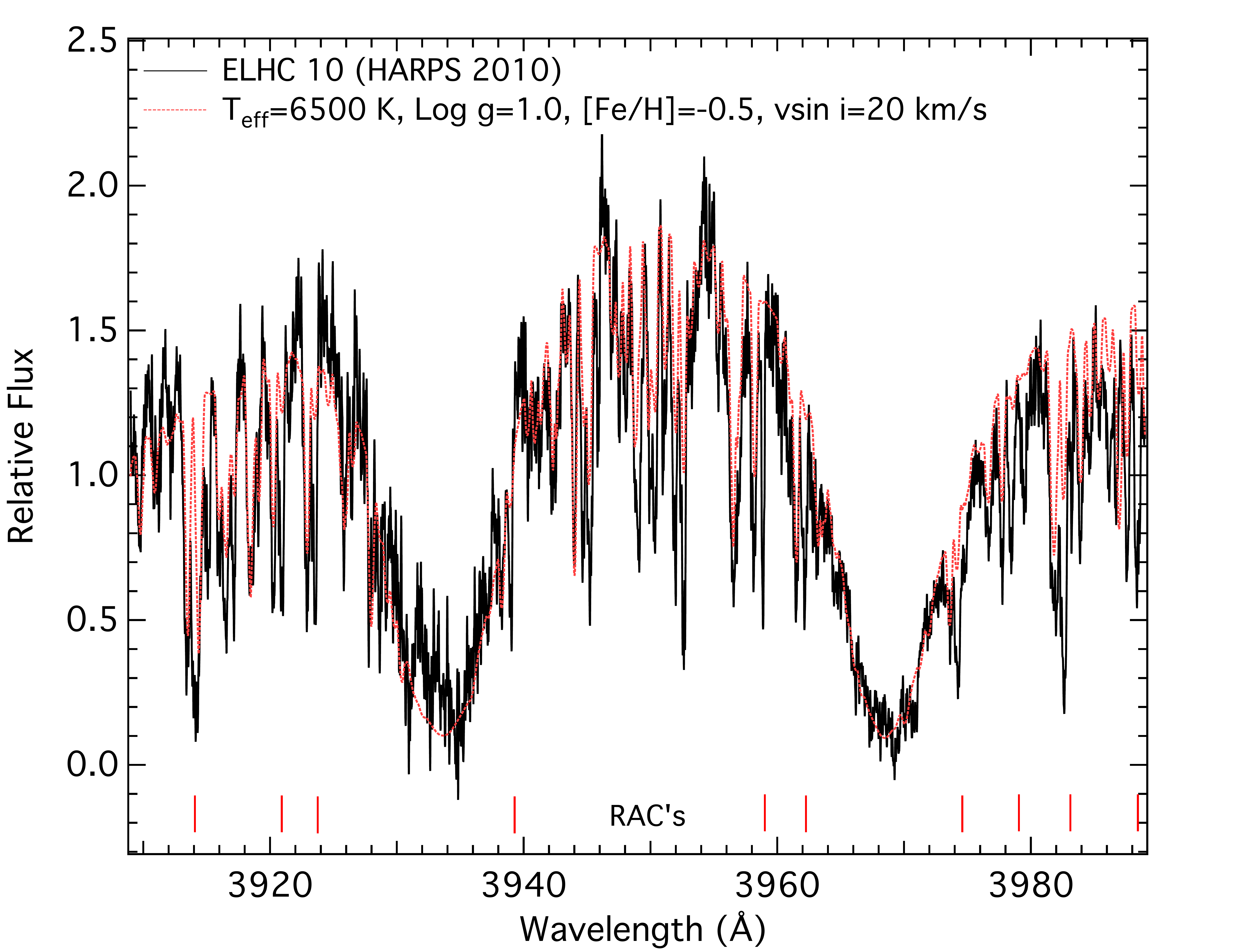}
          \includegraphics[width=0.495\textwidth]{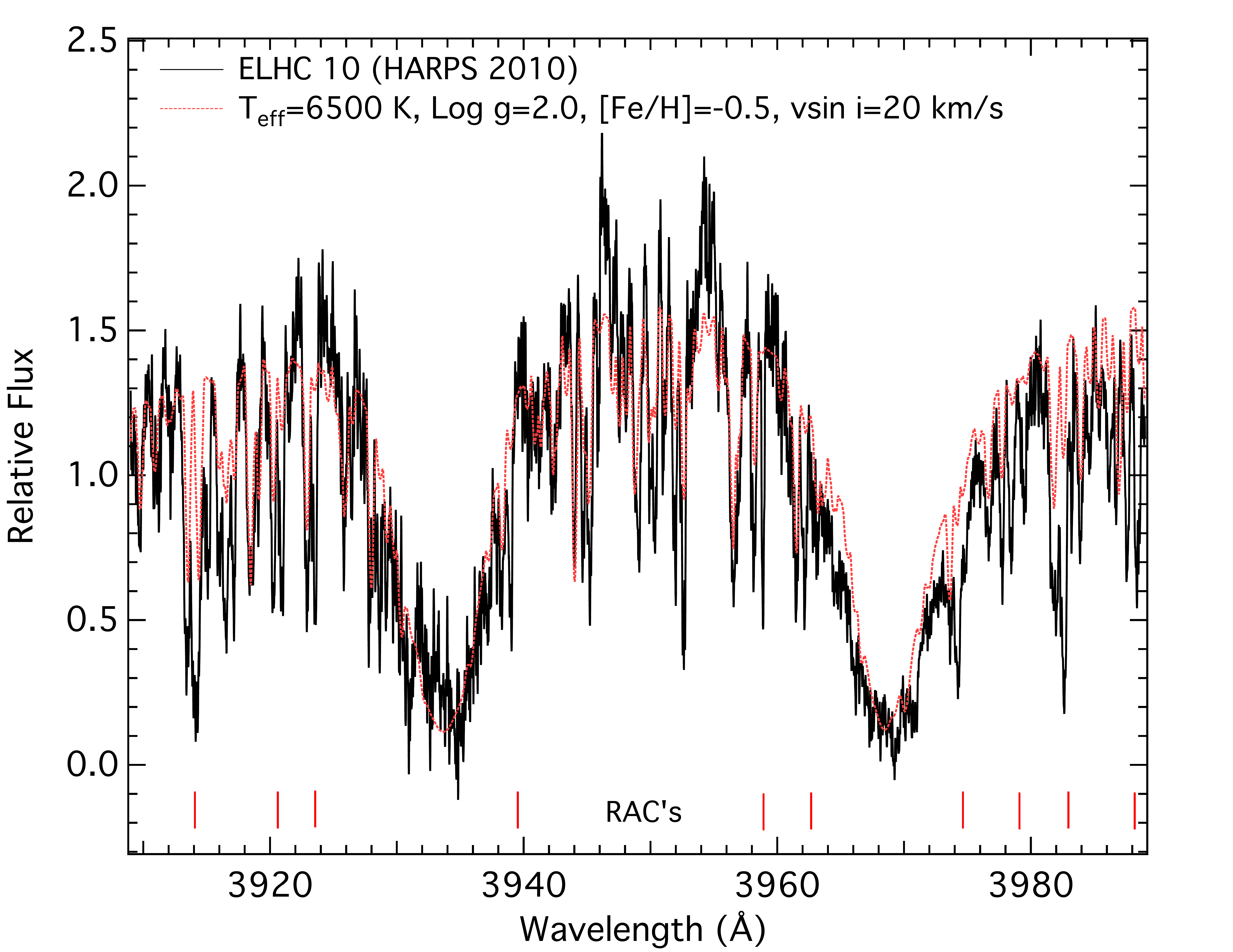}
    \includegraphics[width=0.495\textwidth]{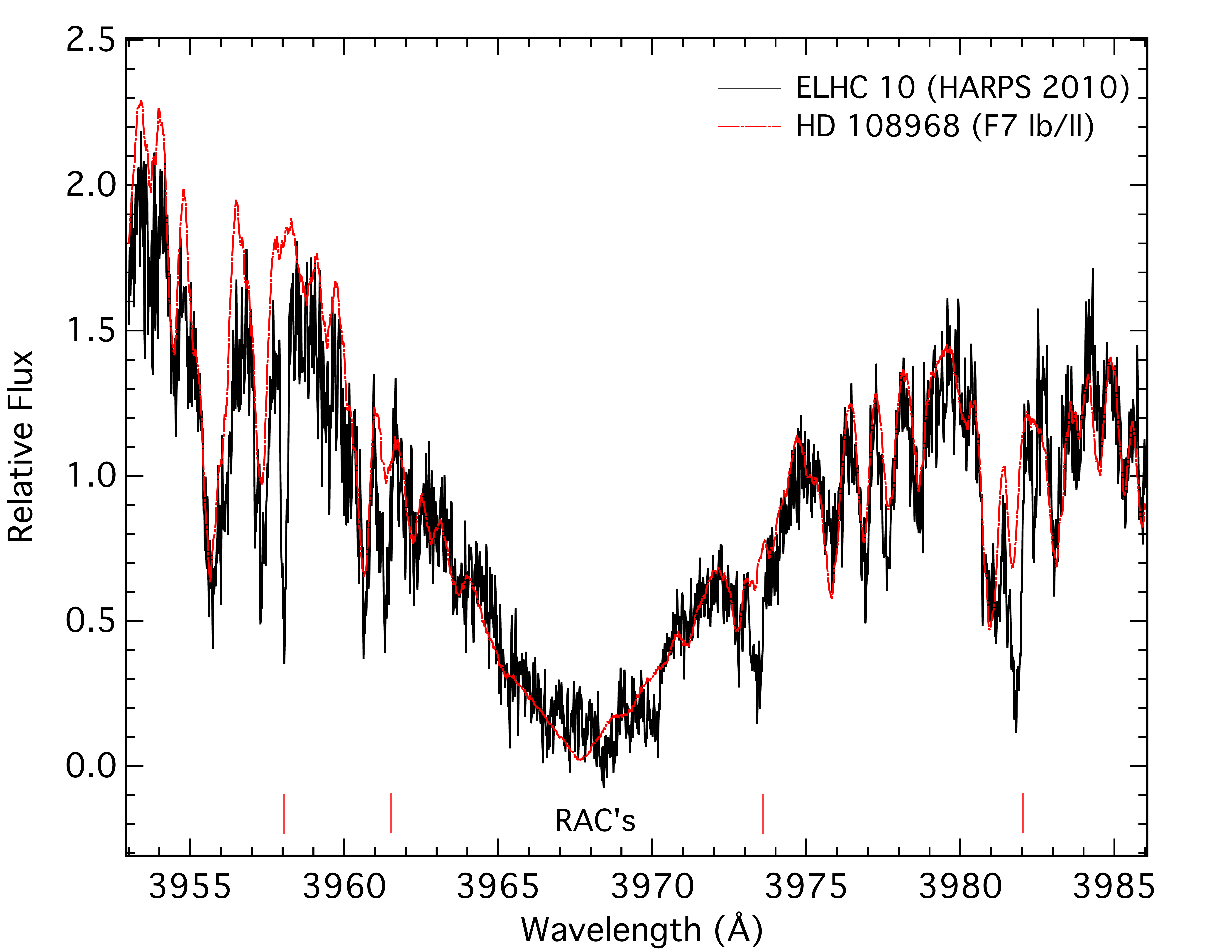}
     \includegraphics[width=0.495\textwidth]{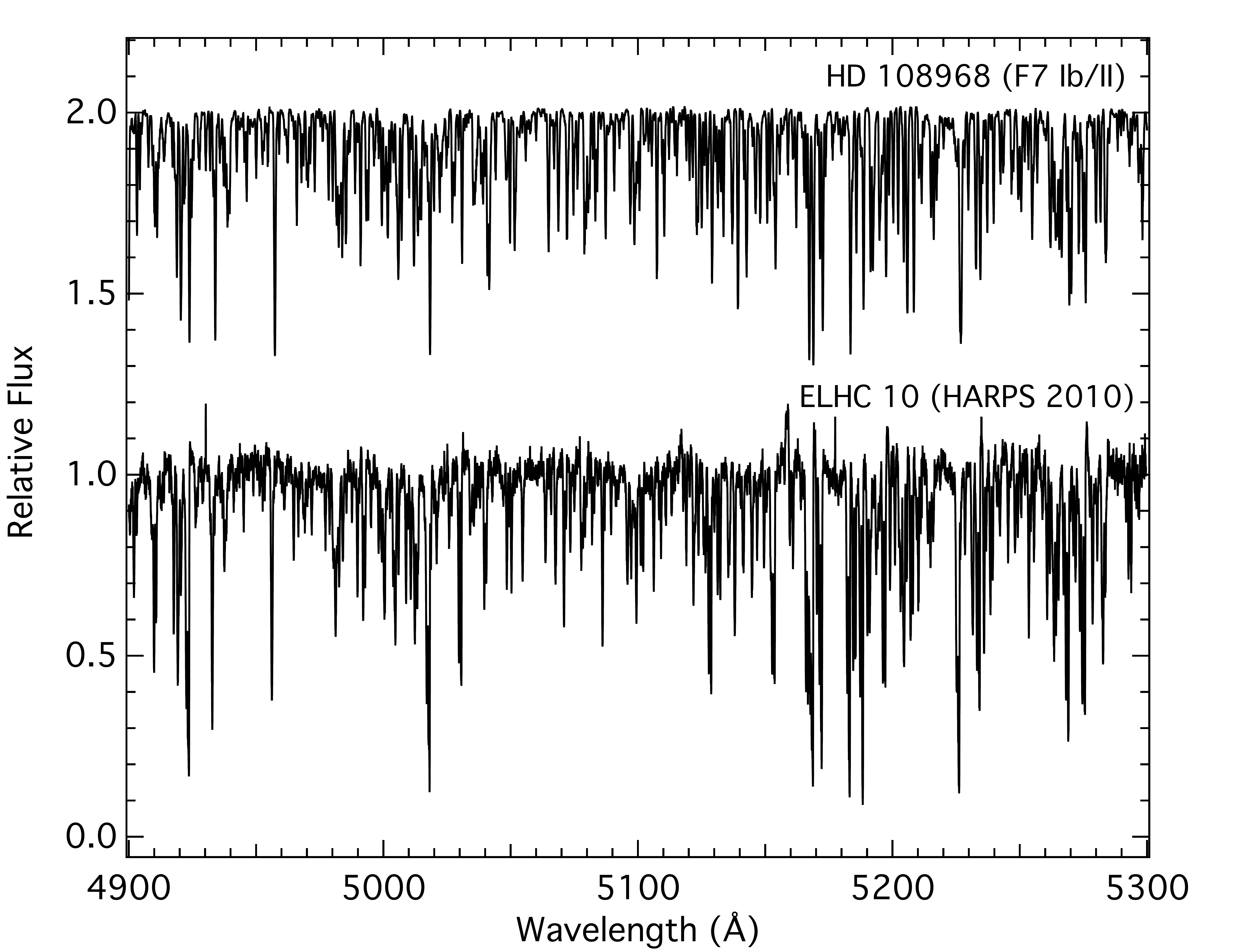}  
    \caption{{\it Upper panel:}  HARPS (2010) spectrum of ELHC\,10  and synthetic spectra from Coelho et al. (2005) with different $\log g$ and  low metallicity.  {\it Bottom panel:}  The optical spectrum of ELHC\,10  and template spectrum taken from UVES atlas covering the H line of Ca {\sc ii}  and metallic lines in the region 4900$-$5300 \AA. Positions of some red absorption components (RACs) are indicated. The template spectrum have been offset in flux for convenience and ELHC\,10 have been blue-shifted to the rest frame. }
    \label{Model}
\end{figure*}

 \begin{figure}\label{SED_ELHC10}
        \includegraphics[width=0.5\textwidth]{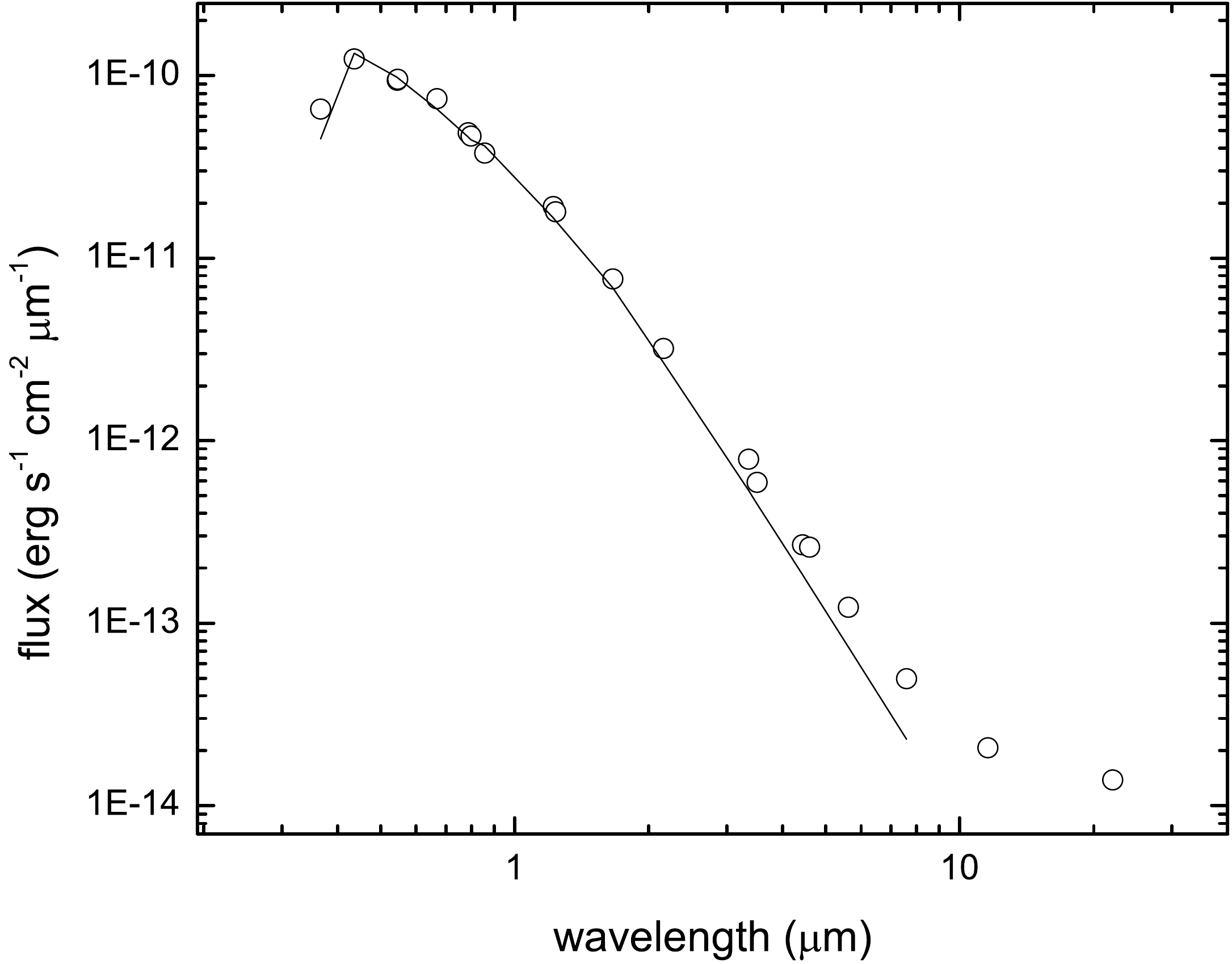}
    \caption{Spectral energy distribution for ELHC\,10, constructed with the  flux values listed in Table\,3 and the best fit.  The two redder fluxes were excluded from the fit since the corresponding W3 and W4 WISE images does not show the presence of the system, only the sky at the target position.}
    \label{SED}
\end{figure}

 \begin{table}\label{Tabla2}
 \centering
\caption{Fluxes and their errors derived from magnitudes reported in different databases.}
\begin{tabular}{ccccc}
\hline
Filter                                               & $\lambda$	&	f$_\lambda$   &$\sigma$\,f$_\lambda$ \\
                                                        & (\AA)	          &	({\small erg/cm$^{2}$/s/\AA})	         &	    ({\small    erg/cm$^{2}$/s/\AA })         \\\hline	
B                                                      &  4413.08	&	1.30E-14	         &	1.79E-16		\\
V                                                     & 5512.12	          &	9.38E-15	         &	1.30E-16		\\
DENIS I & 7862.10	          &	4.97E-15     	&	1.37E-16	\\
I                                                       & 8059.88          &	4.86E-15       	&	2.24E-17	 \\
J                                                      & 12350	          &	1.66E-15	         &	6.11E-17		 \\
H                                                     & 16620	          &	7.48E-16       	&	2.76E-17		\\
K$_s$                                             & 21590	          &	2.91E-16       	&	1.61E-17		\\
WISE1& 33526	          &	7.93E-17     	&	2.05E-18	\\
IRAC1   & 35634	          &	5.79E-17    	&	1.60E-18	\\
IRAC2  & 45110	          &	2.61E-17    	&	6.97E-19	\\
WISE 2 & 46028  	&	2.60E-17    	&	6.46E-19		\\
IRAC3   & 57593	         &	1.17E-17     	&	5.41E-19		\\
IRAC4  & 79594         	&	4.67E-18     	&	7.74E-19		\\
WISE3  & 115608	&	2.08E-18   	&	1.57E-19	\\
WISE4  & 220883	&	1.38E-18   	&	3.50E-19	
\\\hline
\end{tabular}
\vspace{0.2 cm}
\noindent\parbox[c]{8 cm}{ \small{References:} I data from DENIS Deep Near Infrared Survey \citep{Cioni2000}, $BVI$ data from OGLE II - III survey \citep{Udalski2000, Udalski2008},  
$JHK_{s}$ data from 2MASS all-sky survey \citep{Skrutskie2006}, {\it IRAC} \citep{Meixner2006} and {\it WISE } using data retrieved from the {\it Spitzer Space Telescope} Archive \citep{Wright2010}.}
\end{table}

\begin{figure}
     \includegraphics[width=0.495\textwidth]{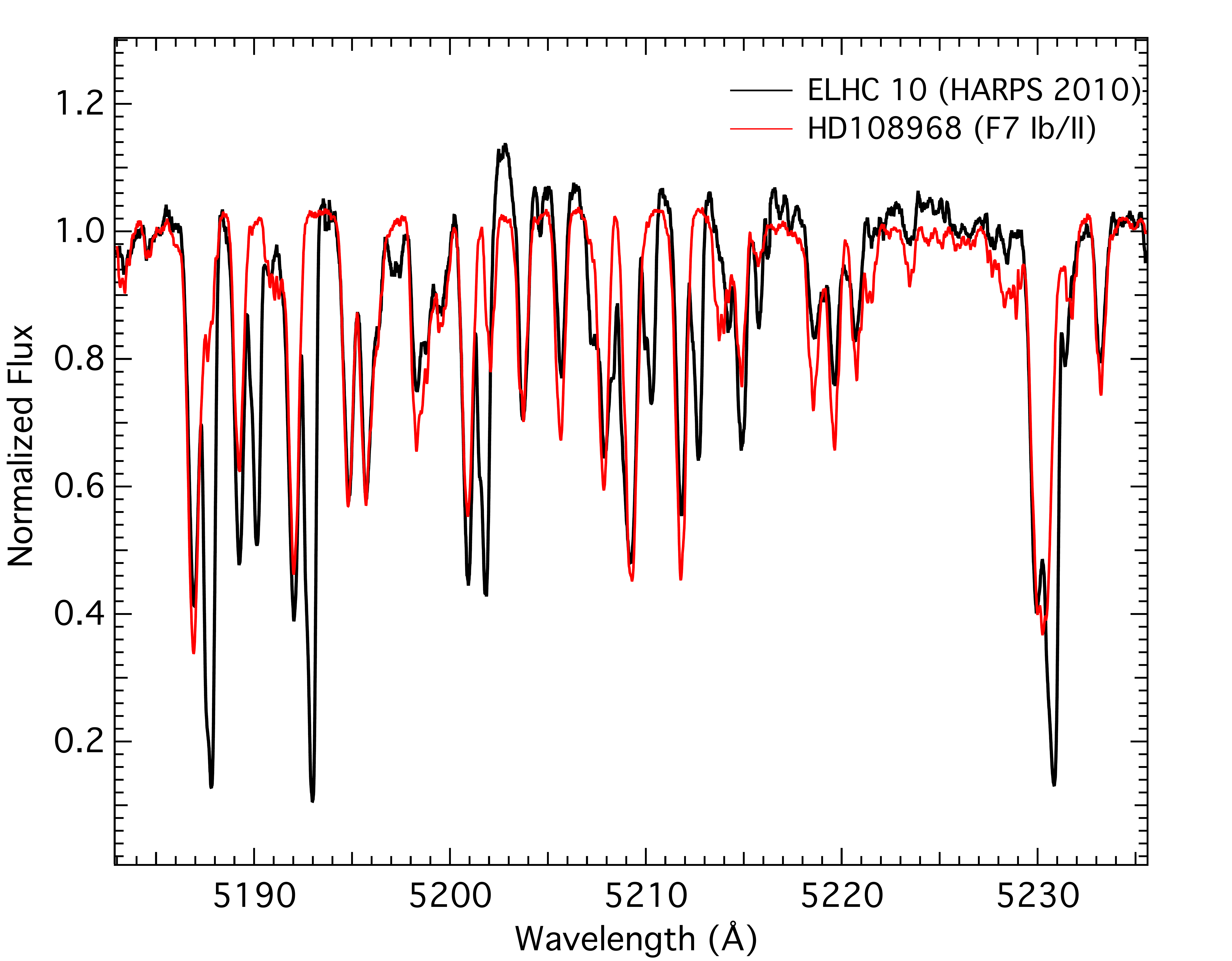}
    \caption{Spectrum of ELHC\,10 overplotted with the template spectrum, exhibiting  red absorption component (RAC's) in the metallic lines.}
    \label{fig4}
\end{figure}

\begin{figure}
    \includegraphics[width=0.5\textwidth]{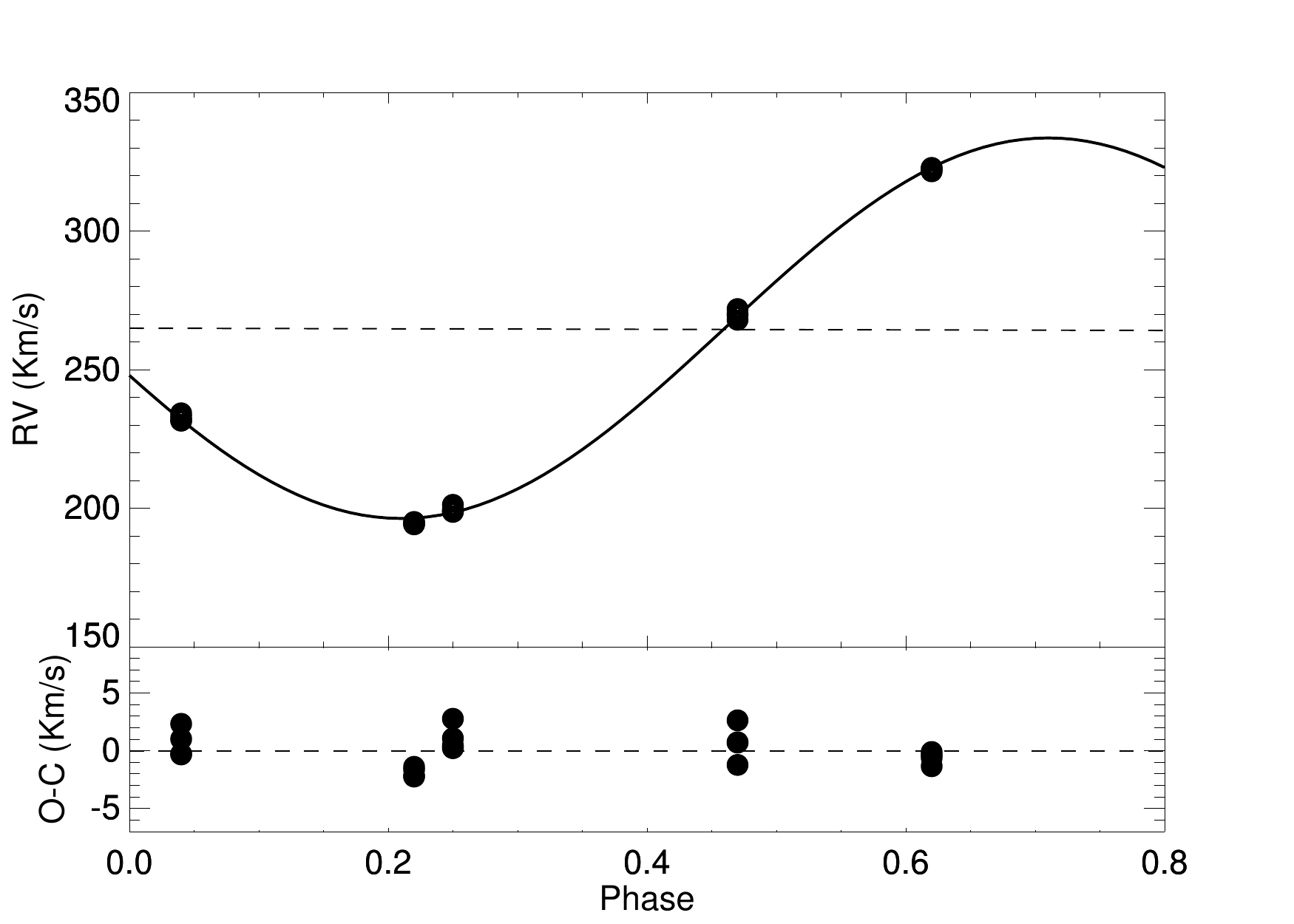}
    \caption{ {\it Upper panel}: radial velocities of  ELHC\,10 phased with a period 219.9 days and the best-fitting solution. The horizontal dashed line marks the corresponding 
    systemic velocity. {\it Bottom panel}: residuals from the fit.}
    \label{RVs}
\end{figure}

\subsection[]{Temperature and surface gravity of the  primary star}\label{SED}

 We compared our spectra with the library of synthetic stellar spectra given by \cite{Coelho2005}. The synthetic spectra with low metallicity were adopted, for which, upper and lower limits of the metallicity were established from the mean values of metallicity distribution derived by \cite{Carrera2008} for the chemical enrichment history of the LMC. The spectra with effective temperatures 
$3500\leq T_{\mathrm{eff}}\leq 7000$ and surface gravities $0.0\leq \log g\leq 3.0$  were subtracted from our spectrum in the region  3911-3993 \AA\  covering the Ca {\sc ii} H/K  lines in order to find  the residual spectrum with the smaller standard deviation, considered as the best model spectrum.

We find that a spectrum with $T_{\mathrm {eff}}=6500$ K, $\log g = 1.0$ and  [Fe/H]$=-0.5$ reproduces well this part of the HARPS spectrum in 2010. We estimate the accuracy of our adjustment to be $\pm\, 250$ K in T$_{eff}$ and $\pm\, 0.5$ dex in $\log g$ and  [Fe/H]. In addition, we used the tabulated effective temperature of supergiants as a function of spectral type  from \cite{Kovtyukh2007} to independently verify the range of spectral type for ELHC\,10 based on its $T_{\mathrm{eff}}$: F3 (6700 K)$-$F6 (6270 K), which is consistent with the photometric estimation. 
In Fig. \ref{Model} we show the HARPS spectrum taken in 2010 for ELHC\,10,  along with the best synthetic spectrum and also the spectrum of the standard HD\,108968 (F7 Ib/II, v\,$\sin i = 22 \pm 1.8$ km/s) taken from the atlas UVES-POP\footnote{http://www.eso.org/sci/observing/tools/uvespop.html} \citep{Bagnulo2003}. 

The fact that no veiling was necessary to fit the lines in the spectrum with a supergiant template indicates that most of the optical flux comes from this star, and the contributions of the companion and circumstellar material are very small at these wavelengths.

\subsection{Analysis of the spectral energy distribution}

We compiled fluxes  at different wavelengths from several sources to build the spectral energy distribution (SED)  from optical to far-infrared. Magnitudes $m_{\lambda}$  
were transformed to fluxes $f_{\lambda}$ using the standard zero magnitude fluxes; the results are shown in Table\,3. 

We  performed a fit to the SED by means of Marquant-Leveberg non-linear least-square algorithm by minimization of $\chi^{2}$ of the function:

\begin{equation}\label{2}
f_{\lambda}=f_{\lambda,\, 0}10^{-0.4E(B-V)[k({\lambda}-V)+R(V)]},
\end{equation}

where
 \begin{equation}\label{3}
f_{\lambda, 0}=\big(R_{1}/d\big)^2\big[f_{1,\lambda}\big],
\end{equation}

\noindent and $f_{1}$ is the flux of the primary star, $k( \lambda-V)\equiv E(\lambda-V)/E(B-V)$ is the normalized extinction curve, $R(V)\equiv A(\lambda)/E(B-V)$  is the ratio of extinction at $V$ to reddening, $d$ is the distance to the binary and $R_{1}$ is the primary physical radius. Since the secondary is not detected in the spectrum
it was not considered in the fit. For further information about the fitting procedure see \cite{Mennickent2010c}. 

The parameters $T_{\mathrm{eff}}$ and $\log g$ obtained in Section 3.2 were used to obtain a synthetic spectrum with metallicity $-2$  from the grid of BT-NextGen (AGSS2009) spectra available at the Spanish Virtual Observatory\footnote{http://svo.cab.inta-csic.es/main/index.php} and the parameters $R_{1}/d$ and $E(B-V)$ were the free parameters. 
We minimized the $\chi^{2}$ between the observed SED and the synthetic spectrum and converted to absolute units using the distance  to the LMC of $\mu_{LMC}= 18.493\pm0.008$ mag \citep{Pietrzynski2013}. 

The result of the fit gives $R_{1}= 61  \pm 10  \,R_{\odot}$ and $E(B-V)= 0.039 \pm 0.089$, compatible with the excess derived from the H$\alpha$ emission strength in Sec.\, 4.4.  We calculate the ranges of primary mass from 
 \begin{equation}\label{primary_mass}
\log g=\log M_{1}-2\log R +4.437,
\end{equation}
where $M_{1}$  and $R$ are in solar units and surface gravity is in CGS system of units.
The radius and surface gravity are compatible with a  primary mass $M_1$ between 0.3 and 5.9 M$_{\odot}$. 
The fit and the data are shown in Fig.\,5; the small infrared excess remaining at WISE $W_{1}$ and $W_{2}$ bands could indicate emission from  circumstellar material. 
The obtained radius excludes the main sequence nature for the primary, as expected from the large brightness of this cool star located in the LMC.

\begin{figure*}
    \includegraphics[width=0.95\textwidth]{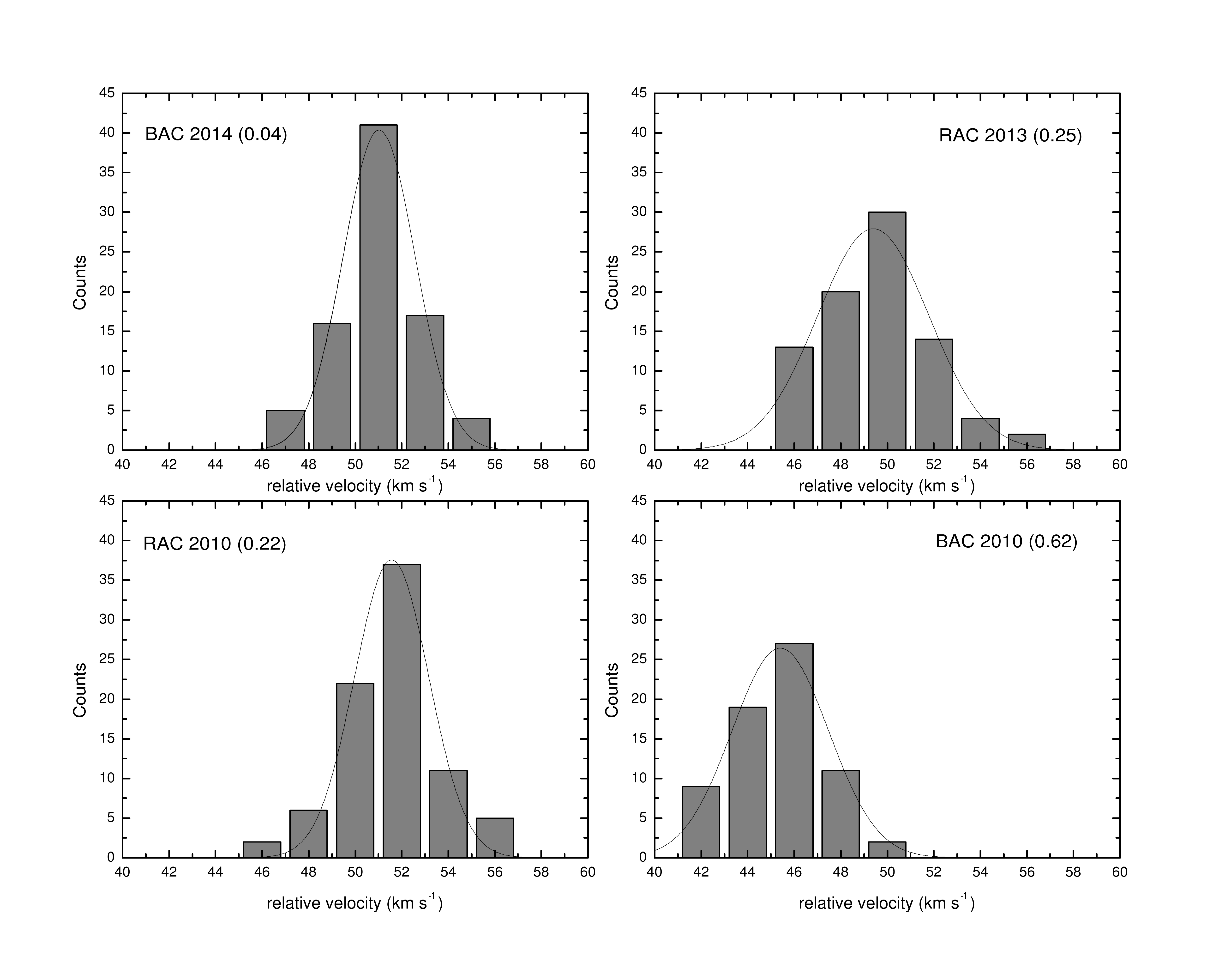} 
    \caption{ Distributions of radial velocities (relative values) of the associated RAC/BAC relative to the stellar line component at different orbital phases. 
    The best gaussian fits to the distributions are also shown. Their parameters are given in Table 6.}
    \label{histogram}
\end{figure*}

  \begin{figure}
    \includegraphics[width=0.5\textwidth]{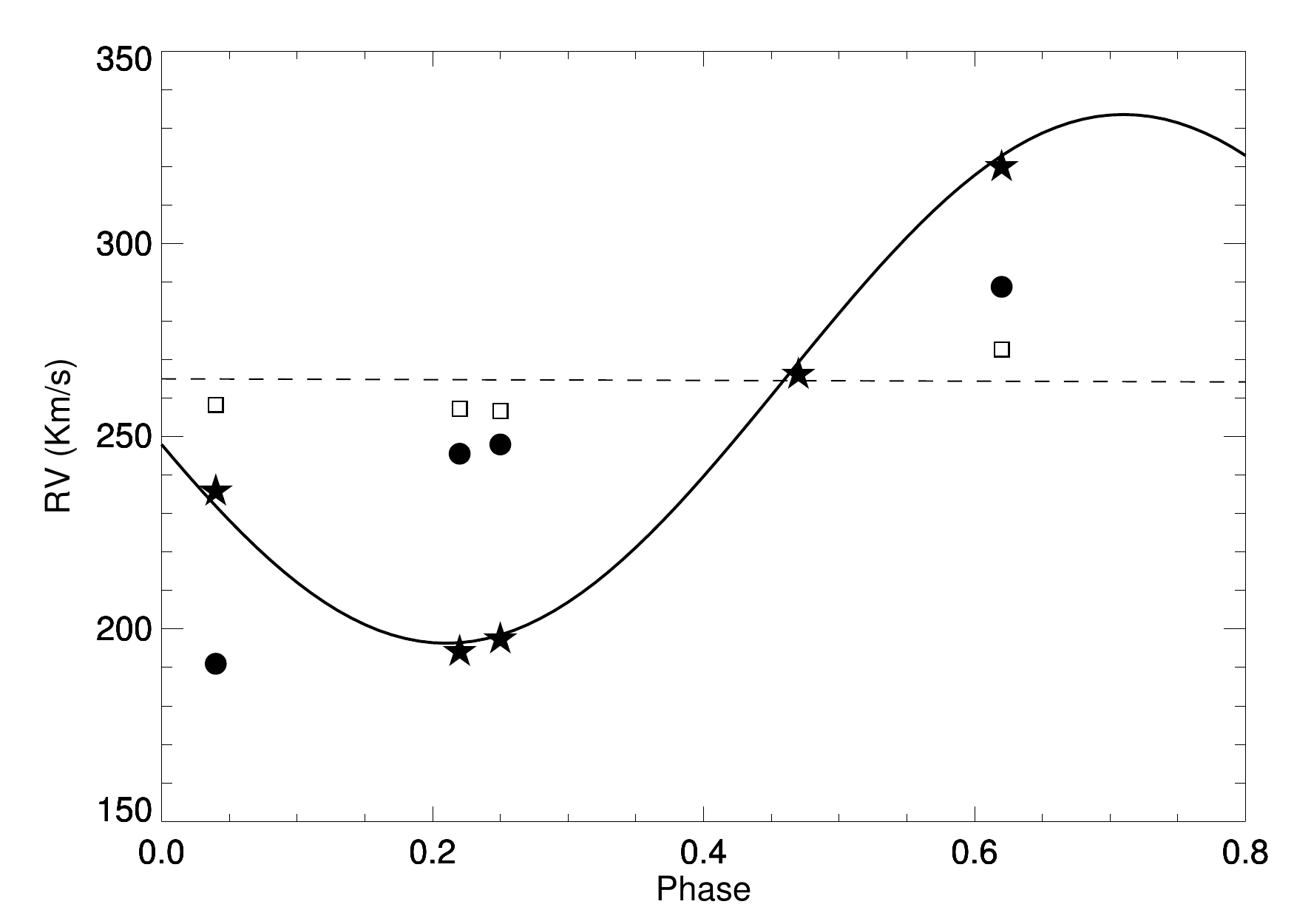}
    \caption{Radial velocities of  Fe {\sc ii} lines at 5018 \AA\, phased with a period 219.9 days and the best-fitting solution derived in Section \ref{photospheric_lines}. 
The horizontal dashed line marks the corresponding systemic velocity, star black points indicate the photospheric
line velocity of the F-supergiant star, black points represent the velocity
of the DACs and squares the velocities of narrow absorption components.}
    \label{RVs_FeII}
\end{figure}

\begin{table*}\label{RVs}
\caption{ Heliocentric radial velocities for ELHC\,10. Residuals from the best fit function are indicated as $O-C$.}
\begin{tabular}{ccccccccc}
\hline
Phase & Y\,{\sc ii}  (4\,883.7 \AA )& $O-C$ & Ba\, {\sc ii} (4\,934.1 \AA) & $O-C$ & Ba\, {\sc ii} (5\,853.7  \AA) & $O-C$ &  Ca\,{\sc i}  (5\,857.5 \AA)& $O-C$\\
          & (km s$^{-1}$) &(km s$^{-1}$)  & (km s$^{-1}$)&  (km s$^{-1}$)& (km s$^{-1}$) & (km s$^{-1}$) & (km s$^{-1}$) & (km s$^{-1}$) \\\hline
 0.04 &  $232.919\pm 0.112$ &  1.016   &  $231.587\pm 0.485$ &  -0.317  & $231.623\pm 0.059$ &  -0.281& $234.234\pm0.219 $ &  2.330\\  
  0.22 &  $195.037\pm 0.010$ &  -1.392 &  $194.793\pm0.004 $ &  -1.636  & $194.252\pm 0.083$ &  -2.176 & $194.193\pm0.523 $ &  -2.236\\  
  0.25 &  $198.663\pm 0.131 $ &  0.221  &  $198.880\pm0.448 $ &  0.438  &  $199.516\pm 1.132$ &  1.074 & $201.210\pm 1.527$ &  2.768\\        
  0.47 &  $269.915\pm0.117 $ &  0.711  &  $267.977\pm 0.028$ &  -1.228 &  $269.933\pm 0.052 $ &  0.729 & $271.836\pm0.566 $ &  2.632\\        
  0.62 &  $322.264\pm 0.187 $ & -0.604 &  $321.525\pm 1.301$ &  -1.342 &  $322.760\pm 0.049$ &  -0.108 & $322.510\pm0.220 $ &  -0.358\\                    
\hline
\end{tabular}
\end{table*}

\subsection[]{Radial velocity study of photospheric lines}\label{photospheric_lines}

The optical and near-infrared spectrum of ELHC\,10  is dominated by metallic lines of the F-type primary. The spectral lines are redshifted by amounts of roughly 3 \AA, typical for LMC stars. Interesting, 
some metallic lines are split in discrete absorption components  (Fig.\,6). 
Line identification was made by comparing line positions with theoretical line wavelengths \citep{Kurucz1993}. Radial velocities (RVs) were measured by calculating the positions of the line centre with a 
Gaussian fit using the IRAF {\it SPLOT} task. In the case where lines were blended, we employed a deblending method to isolate the components  by fitting multiple gaussians to the observed profiles.
We performed this task with the {\it IRAF} deblending `d' routine available in the  {\it SPLOT} package.

Firstly, we measured RVs of  lines clearly representative of the primary motion. The four absorption lines used for this purpose were Ba\, {\sc ii} 4934.1 and 5853.7 \AA, Ca\,{\sc i} 5857.5 \AA\ and Y\,{\sc ii} 4883.7\AA.
These lines were chosen because they are ``well behaved", i.e. they are not too weak to be measured, they are not contaminated by telluric features, they are not blends with lines from other elements, the lines are 
almost always symmetrical in comparison with the other metallic lines and they do not show additional discrete absorption components  similar to other metallic lines in the optical spectra of ELHC\,10. 
In Table\,4 we report the RVs determined for each of these lines. The measures reflect very well the orbital motion of the primary (Fig.\,7).

We assume a circular orbit and fit the  RVs with a sine function of the form:

\begin{equation}\label{5}
RV = \gamma + K sin\, \big(2 \pi\,(\Phi + \Phi_{0}) \big).
\end{equation}

\noindent
Considering all velocities with equal statistical weight we  find a system velocity $\gamma_{\mathrm{1}}=264.935\pm 0.432$ km\,s$^{-1}$, semi-amplitude $K_{\mathrm{1}}=68.639\pm 0.563$ km\,s$^{-1}$ 
with rms 6 km\,s$^{-1}$ and phase shift $\Phi_{0}=-0.460\pm 0.01$.  The existence of a small phase shift between the spectroscopic and photometric ephemerides is compatible with the fact that we use the ephemerides for the very minimum of the light curve, which is shifted by about $+0.05$ from the minimum of the light curve assuming symmetrical eclipse.
We reviewed the spectra throughout the spectral range to find  that there is no indication for the secondary component even at minimum. 

\subsection{Radial velocity study of non-photospheric lines}

The other prominent lines seen in the spectra are due to Ca\,{\sc ii}, Na\,{\sc i} and a large number of lines from Fe\,{\sc ii}, Ti\,{\sc ii}, Na\,{\sc i} and Si {\sc ii} and other species. 
They display  line splitting as discrete red or blue extra absorptions depending on the phase in which the spectrum was obtained. These DACs were  denominated RACs or BACs; they are not due to 
another stellar component   because their variable morphology is incompatible with this assumption. Once the orbital motion of the primary was determined, DACs were clearly identified in the 
lines. In this section we report the RV measurements of these DACs.

 \begin{figure}
    \includegraphics[width=0.5\textwidth]{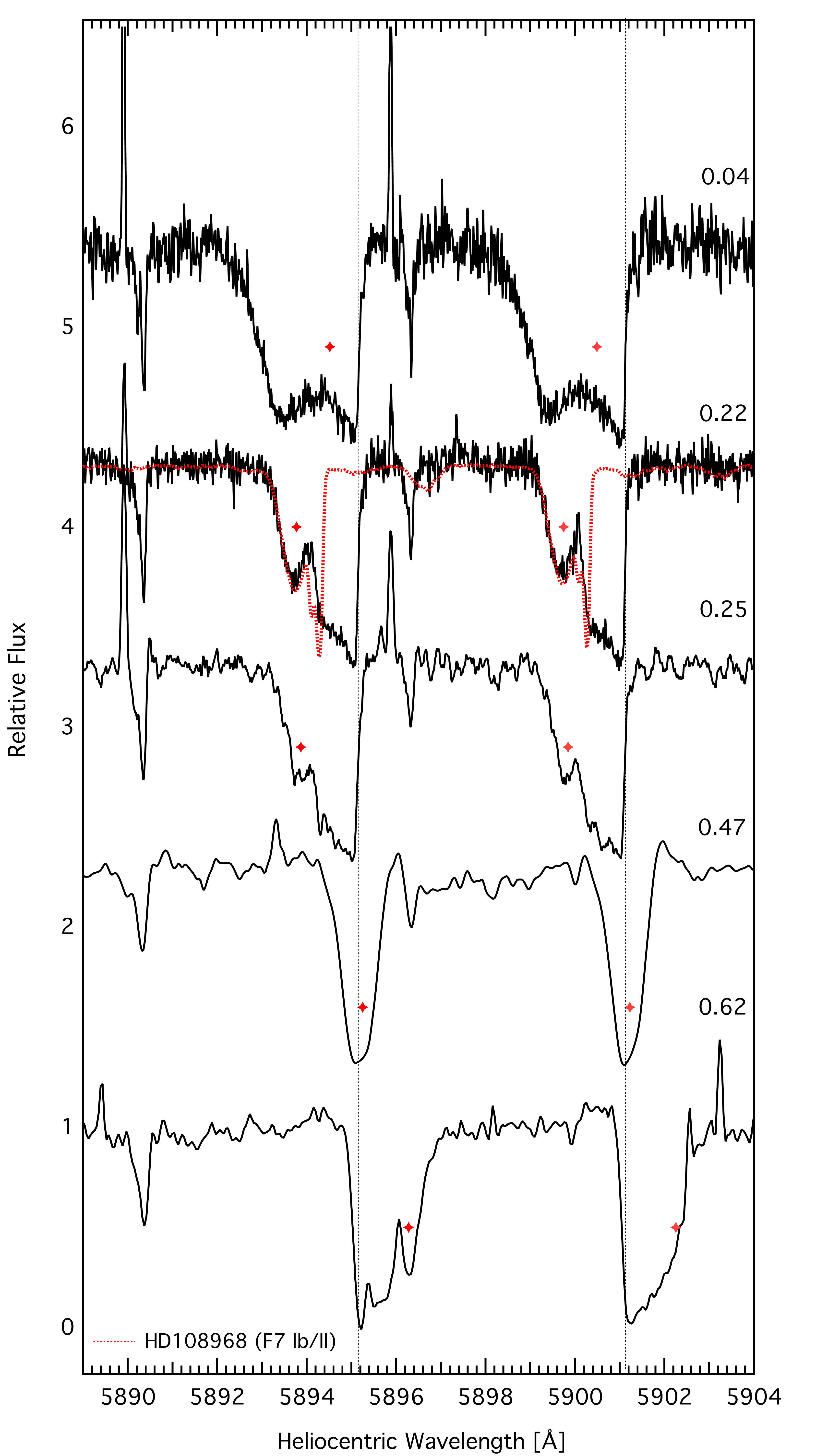}
    \caption{Na\,{\sc i }\,D\,{\small1} and D\,{\small2} lines  as a function of the heliocentric wavelength. 
    The dashed lines mark  our systemic velocity of 265 km\,s$^{-1}$ for each line, the red points indicate the photospheric velocity of the F-supergiant star. 
    The spectrum of the  standard HD 108968 is overplotted as illustration at phase 0.22. It is possible to observe the separated contributions due to stellar and circumstellar media in 5\,890 \AA\ and 5\,896 \AA. 
    The labels in the upper right of each spectrum indicate the orbital phase.}
    \label{NaI}
\end{figure}

For the present study we selected 202 prominent lines to obtain accurate radial velocities. The result of this procedure is summarized in Table 5.  We give the average RV of the primary, 
along with the velocity of the additional components for each ion. The internal error, measured by weighting the internal $rms$ of individual ions in photospheric lines,
usually do not exceed $\pm\,3\  \mathrm{km\,s}^{-1}$. 
In Fig. \ref{histogram}, we show the distribution of relative velocity between the main (photospheric) component ($\mathrm{V_{Main}})$ and its associated RAC or BAC ($\mathrm{V_{Rac/Bac}}$).  

We performed a non-linear least squares fit to the DAC velocity distributions using a gaussian function of the form:

\begin{equation}
 y = A \exp \Bigg(-\frac{(x-x_{c})^{2}}{2w^{2}} \Bigg).
 \label{6}
\end{equation}

\noindent The coefficients $A$, $x_{c}$ and $w$ are given in Table 6. We notice relatively narrow distributions and that the velocity of blue or redshifted absorptions are always near $\pm\,50$ km\,s$^{-1}$, except at 
$\Phi_{o}$ = 0.62 when they are around  $-45$ km\,s$^{-1}$ (Fig.\,8). 

We also find narrow absorption components at velocity 252 km s$^{-1}$ at the Si\,{\sc ii} 6347 and 6371 \AA\, lines. These absorptions are visible only at main eclipse, they have $FWHM$ $\sim$ 3  km\,s$^{-1}$  and 
are shifted by $-8$ km\,s$^{-1}$   with respect to the velocity of the system center of
mass; they might be produced in a slowly expanding disc surrounding
the binary or in a wind  perpendicular to the orbital plane.
 We discard the origin in the supergiant because of the very narrow
line width, not comparable with the photospheric lines seen in the spectrum. The lines are also too narrow to be formed in any Keplerian disc rotating near the stars.

As an example of radial velocity for DACs we show in Fig.\,\ref{RVs_FeII} the case of Fe\,{\sc ii}\,5018 \AA. It is clear how the photospheric line follows the orbital motion whereas the
DAC moves from blue to red and the velocity of the narrow absorption component remains stable.

\begin{table*}\label{Table4}
\caption{Summary of heliocentric radial velocities (in km s$^{-1}$) for the spectra in seasons 2010, 2012, 2013 and 2014. We give the velocity of the main component followed by the velocity of 
the associated RAC\,/\,BAC relative to this component. The number of lines used in the averages is listed between parentheses. Errors reflect the rms of the RVs per lines within an ion. 
Note that at phase 0.47 (season 2012) no RAC\,/\,BAC were observed. }
\begin{tabular}{cccccc}
\hline
Ion	                    &	2014 (BAC)                                             &2010 (RAC)	                   & 2013 (RAC)                 &	   2012  	                                                        & 2010 (BAC)\\
                             &   $\Phi_{0}=0.04$                                     &$\Phi_{0}=0.22$                       & $\Phi_{0}=0.25$             &  $\Phi_{0}=0.47$                                           &$\Phi_{0}=0.62$      \\\hline
Ba II                     &	$  232\pm\,1\,(4); 51\pm\,2\,(3) $           &$195\pm1\, (4)$                         & $199 \pm	1 \,(4)$            &  $ 269\pm 1\, (4) $                                   &  $ 321 \pm\,1 \,(4)$          \\ 
Ca I                      &	$  234\pm\,3\,  (3); 47\,(1) 	$                   &$198\pm3\, (3);53\,(1)$                & $199	\pm 3\,(3); 55 \,(1)$        &  $ 271\pm 3\,  (3) $                          & $ 324 \pm\,2 \,(3); 46\, ( 1) $     \\
Ca II                     &	$  220\pm\,1\,  (2)	  $                         &$210\pm 19\, (2)$                        & $217	\pm 29\,(2)$          &  $ 271\pm 15\, (2)$                                   &      \ldots      \\
Cr II                      &	$  232\pm\,1\,  (9);50\,\pm\,2 \,(4)$        &$195\pm2\,(9);50\, (1)$                    & $	198	\pm 1 (9); 46\, (1)$      &  $ 268\pm 3 \,(9) $                             & $  321\pm\,2 \,(9); 46 \,( 1) $           \\
Fe I                      &	$  232\pm\,1\;  (20); 50\,\pm\,3 \,(12) $  &$195\pm 2\,(20);51\pm 2\,(14)    $   & $	198	\pm 2(20); 49\pm 3\,(14) $   & $ 267\pm 2 \,(20) $               & $ 322 \pm\,2 \,(20); 47 \pm\,2 \,( 10) $    \\
Fe II                     &	$  233\pm\,2\;  (26);50\,\pm\,2 \,(26)$    &$194\pm 2\, (28);51\pm 2 \,(28)  $    & $	198	\pm 2(28); 49\pm 3\, (28)$  & $ 267\pm 2  \,(28)$              &$ 323 \pm\, 2\,(26);  45\pm\,2 \,(26 ) $      \\
Mg I                     &	$  234\pm\,1\;  (5); 51\,\pm\,1 \,(3)$      &$195\pm 2\,(5);49\pm 1\,(3)        $      & $198 \pm 2(5); 47\pm 1\, (3)$  & $ 267\pm 1 \, (5)$                      &$ 323 \pm\,1 \,(5); 44 \pm\, 1\,( 3) $    \\
Mg II                    &	$   231\,(1)$                                           &$195\pm 2 \, (2) $	                     & $196\pm 1\, (2) $   & $ 265\pm 4\, (2)  $                                          & $  321\,(1)$\\
Na I                     &	$  233\pm\, 1\; (2);51\,\pm\,1 \,(2) $     &$192\pm 1\,(2);55\pm 1\,(2) $	    & $198\pm 1\,(2); 49\pm 1\,(2) $   & $ 266\pm 1 \,(2) $                       & $  323\pm\,1 \,(2);  45\pm\,2 \,(2 ) $   \\
Sc I                     &	$   234\,(1); 53\,(1)$                             &$196\,(1);51\,(1)$                               & $ 200\,(1);	50\,(1) $     & $ 267\,(1)$                                                &$ 324 \,(1);  42 \,( 1) $     \\
Sc II                    &	$   234\,(1); 53\,(1) $                            &$196\,(1);52\,(1)$                               & $ 198\,(1);	48\,(1)$   & $ 266\,(1)$                                                  &$  323 \,(1);  41 \,(1 ) $     \\
Si II                     &	$  233\pm\,1\,  (6);52\,\pm\,1 \,(3)$      &$194\pm 4\,(6);45\, (1) $	            & $197	\pm 4\,(6); 41\, (1) $    & $ 268\pm 3\, (6) $                          & $ 321 \pm\, 1\,(6);  44\pm\,1 \,(1 ) $     \\
Sr II                     &	$  235\pm\,1\,  (2);52\,\pm\,3 \,(2)$      &$196\pm 4\,(2);51\pm 6\,(2) $	      & $	199	\pm 3\,(2); 49\pm 7\,(2) $ & $ 268\pm 3\, (2) $                 & $  324\pm\,1 \,(2);  42\pm\,2 \,( 2) $    \\
Ti I                       &	$   233\; (3);51\,\pm\,2 \,(3) $               &$196\pm 1\,(3);51\pm 2\, (3)$	    & $	198	\pm 1\,(3); 49\pm 1 \,(3)$ & $ 269\pm 1\,(3)  $                 & $  322\pm\,1 \,(3);  46\pm\,1 \,(3 ) $   \\
Ti II                      &	$  233\pm\, 1\, (19);51\,\pm\,2 \,(19)$  &$195\pm 3\,(24);51\pm 1 \,(23)$	  & $	199	\pm 4\,(24); 49\pm 2 \,(23)$  & $ 267\pm 2\,  (24)$               &$  323\pm\,1 \,(19); 45 \pm\,2 \,(19 ) $      \\
Y II                       &	$ 232 \pm\, 1\,(6);51\,\pm\,2 \,(3)$      &$194\pm 2\,(6);51\pm 1\, (2) $	   & $	196	\pm 2\,(6); 50\pm 2\, (2)$  & $ 267\pm 3\, (6) $                 & $ 323 \pm\,1 \,(6);  46\pm\,1 \,(1 ) $      \\
Mean (km s$^{-1}$)  &    $233\pm	3;51\pm 2 $                &	$195\pm	4;51\pm 2 $	              & $	198	\pm 5; 49\pm 3 $     & $ 267\pm 2  $                               &  $323\pm 1;45\pm 1 $  \\
Weighted mean (km s$^{-1}$) &  $231; 51$ & $193;50 $	& $197; 48 $ & $	  266	  $ & $323;45$    \\\hline
\end{tabular}
\end{table*}

\begin{table}\label{Gaussian_fit}
\caption{ Coefficient of the Gaussian fits (Eq.\,6) shown in Fig.\,8. }
\begin{tabular}{cccc}
\hline
Phase &$x_{c} $ & $w$ & $A$\\ \hline
0.04 & $51.028\pm0.159$ & $1.547\pm 0.138$ & $40.360	\pm3.300$ \\
0.22 & $51.568	\pm0.140$ &$1.598\pm0.137 $  & $37.564\pm2.775 $ \\
0.25& $49.398	\pm 0.338$&$2.396\pm0.362$ & $27.915\pm3.405$\\
0.62 & $45.401\pm 0.206$ & $2.074	\pm0.211$  & $26.432\pm2.279$ \\
\hline
\end{tabular}
\end{table}

\subsection{Na\,{\sc i} D-line profiles}\label{Na-D}

Perhaps the most interesting parts of the Na\,{\sc i} D-line profiles are the discrete absorption components 
present at almost all orbital phases, appearing as red absorption components  near quadrature ($\Phi=0.22$), 
disappearing near $\Phi=0.47$ and changing to blue absorption components from $\Phi=0.62$ to inferior conjunction  ($\Phi=0.04$, Fig. \ref{NaI}). 
DACs in metallic lines follow the same behavior, and are generally of comparable strength and sometimes stronger than the main component.  
A triple Gaussian fit to the NaD lines at phases 0.22 and 0.25  gives a ratio of equivalent widths close to unity. 
The ratio of the sodium lines is equal to 2 for optically thin material and the ratio approaches unity for optically thick gas.
Therefore the gas  producing the NaD absorption must be optically thick. Because our star is located in the LMC, 
an important contribution of interstellar medium to the sodium lines would not be expected, due to the low metallicity. 
From the above we deduce that the material contributing to the extra absorption components in the  
NaD line comes from circumstellar  material of  high optical depth  in the binary system. 

Such behavior in the shape of the spectral features in  Na\,{\sc i}  lines is similarly observed in the Fe {\sc ii} and Ti {\sc ii} lines.   
But quite different are the changes in the shape of ions such as Ba\,{\sc ii}, Mg\,{\sc ii} and Ca\,{\sc i-ii} which usually do not show DACs (Fig.  \ref{FeBa_II}).

 \subsection{Fe {\sc i} and Fe  {\sc ii}  profiles } \label{Iron_line}

The variations in the profiles in both Fe {\sc ii} 4923 and 5018  lines are
hard to interpret in terms of simple changes in the stellar atmospheric
parameters, such as the effective temperature, surface gravity
and microturbulent velocity (Fig.\, \ref{fig6}). 
Actually, multiple components in
the iron lines, up to 4 components, can sometimes appear.
These components are characterized by different line widths, 
revealing different physical conditions in the formation regions. 
A similar behavior (unexplained)  is reported in the interacting binary candidate
 \vb\, which harbors an A-type supergiant \citep{Mennickent2010b}.
In Fig. \ref{RVs_FeII} we show the radial velocities of 3 components of Fe
{\sc ii} line at 5018 \AA. It is clear how the photospheric line follows the primary motion,
narrow stationary components remain stable with the velocity of the system center of mass,
and DACs move from the blue to the red side of the photospheric
line. A similar behavior is followed by DACs in other metallic lines.

\begin{figure}
    \includegraphics[width=0.5\textwidth]{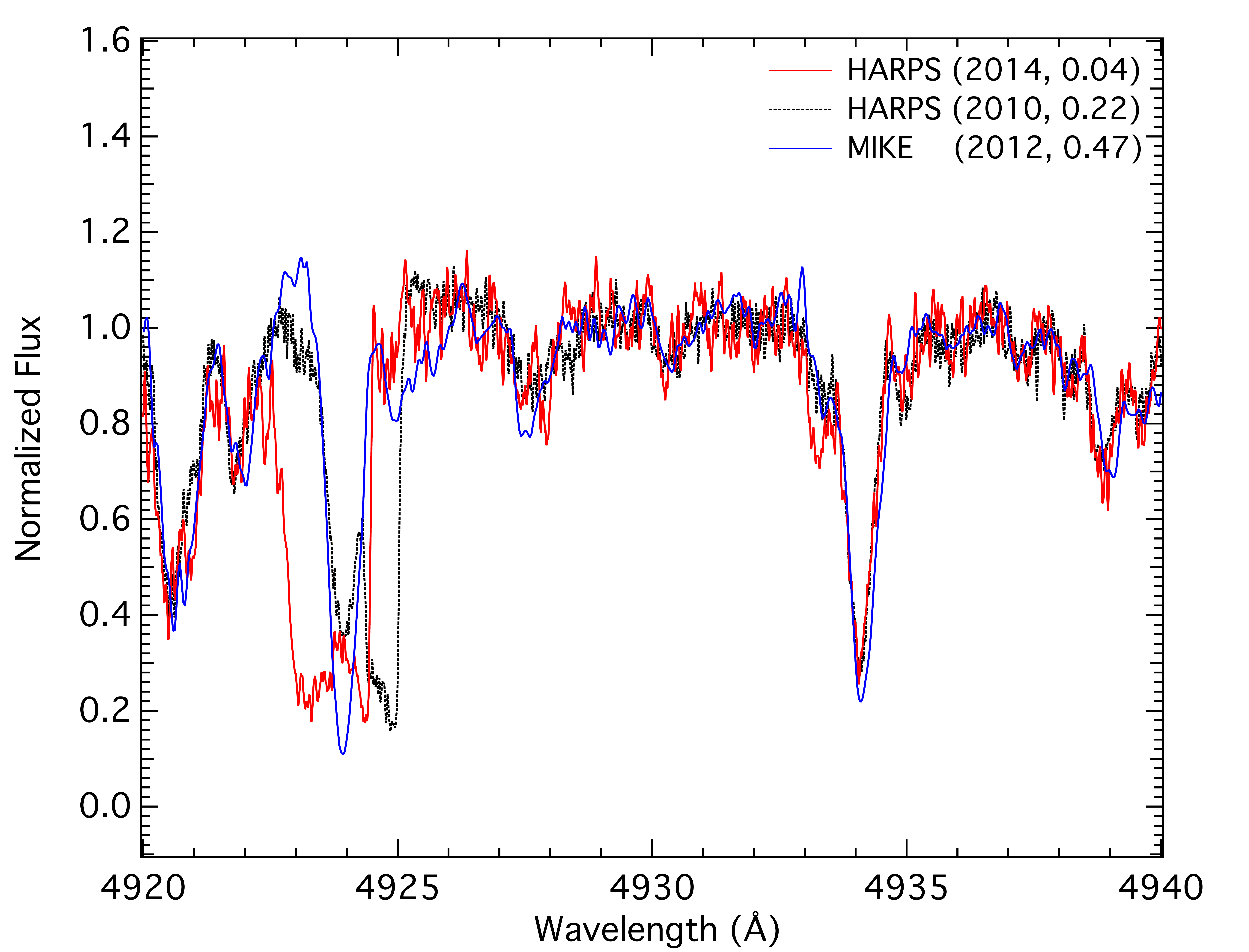}
    \caption{The Fe\,{\sc ii} 4923 \AA\ and Ba\,{\sc ii} 4934 \AA\ lines show the apparition of RACs and BACs and drastic changes with orbital phase in Fe\,{\sc ii}  4923 \AA, 
    whereas Ba\,{\sc ii}  4934  \AA\ is less affected by circumstellar material. Note the almost constant line strength independent of the phase. The spectra are shown in the primary velocity frame. }
    \label{FeBa_II}
\end{figure}

\begin{figure}
    \includegraphics[width=0.5\textwidth]{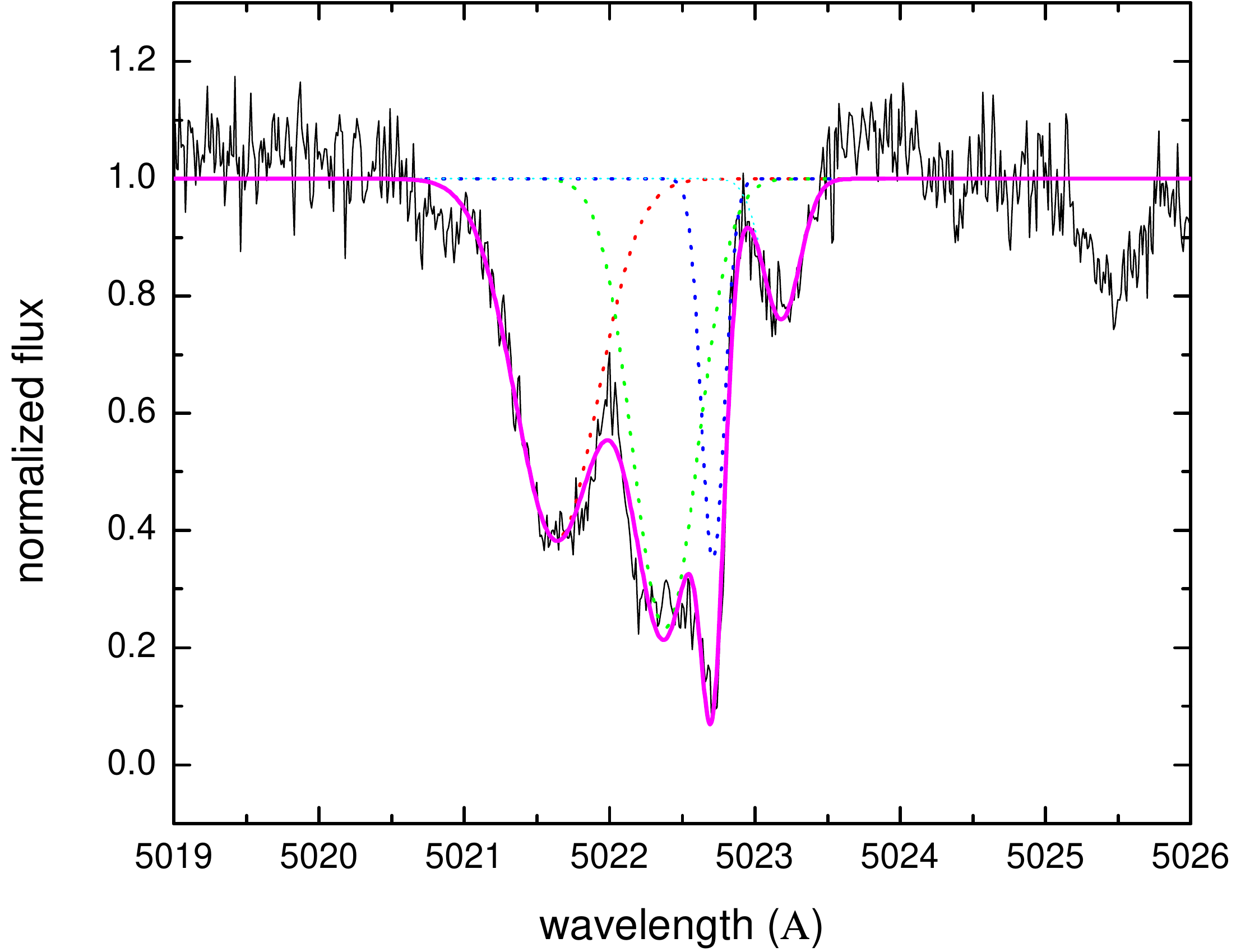}
    \includegraphics[width=0.5\textwidth]{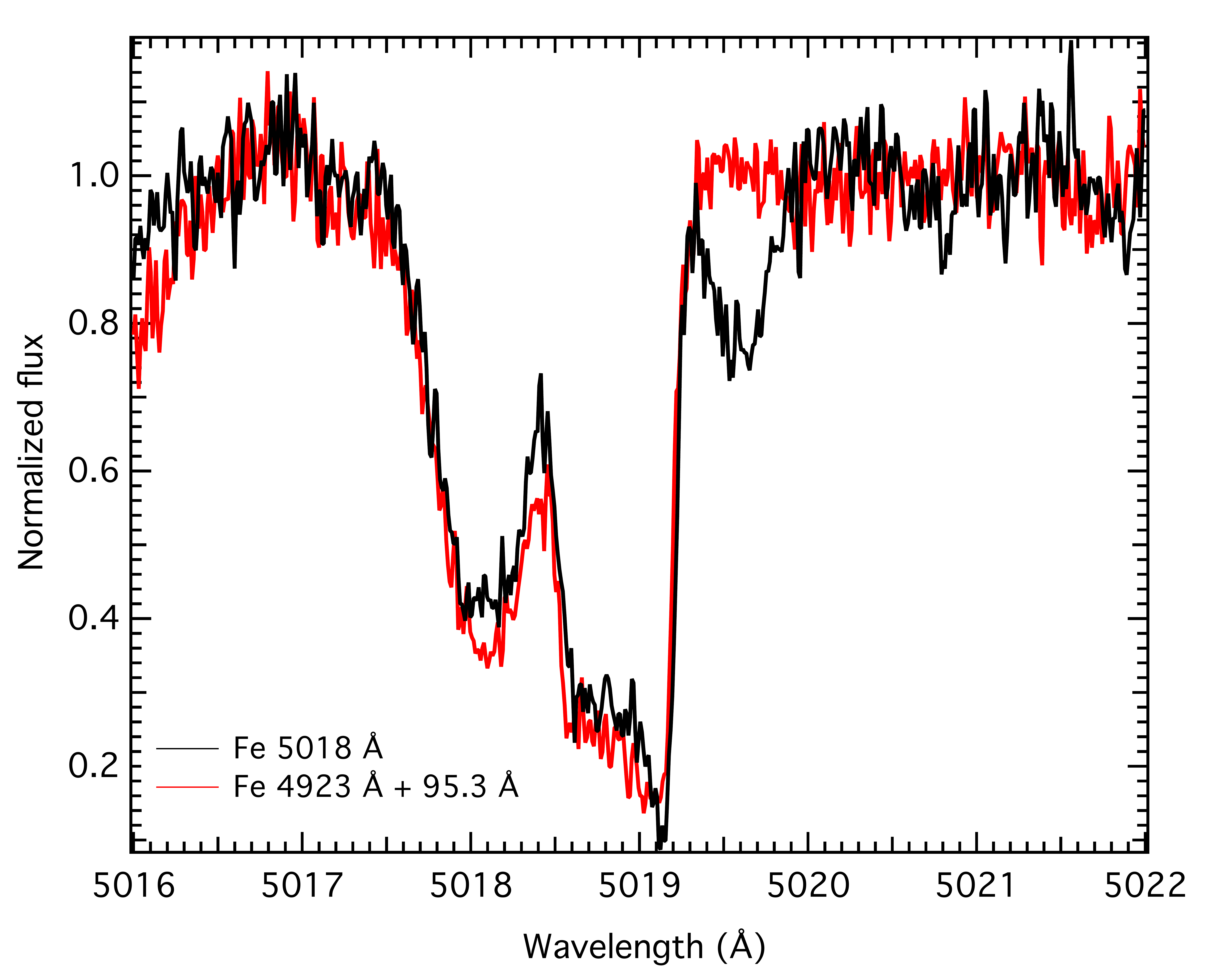}
    \caption{{\it Upper panel:} Multiple Gaussian fit of the Fe\,{\sc ii} 5018 \AA\ line revealing several discrete absorption components. {\it Bottom panel:}  Overplotting of the Fe\,{\sc ii} 4923 \AA\  and 5018 \AA\  lines showing one additional red absorption component in Fe\,{\sc ii} 5018 \AA. The 4953 \AA\  line was redshifted by 95.3 \AA. Both panels refer to the 2010 HARPS spectrum at $\Phi_{o}$ = 0.22. }
    \label{fig6}
\end{figure}

The apparition of these variable and complex double-bottomed shapes in the profile of metallic lines in iron-groups elements, probably reflects large-scale motions with different  radial velocities,
in shock regions, possible representing multiples layers in between the stars. In addition, these spectral characteristics appear and are stronger at or near quadratures and disappear during secondary eclipse, 
as evident in the 2012 spectrum in Figs. \ref{NaI} and \ref{FeBa_II}. All this behavior of features strongly modulated with the orbital period suggests that DACs are produced in gas streams inside the  binary system. 

\begin{figure*}
    \includegraphics[width=1.0\textwidth]{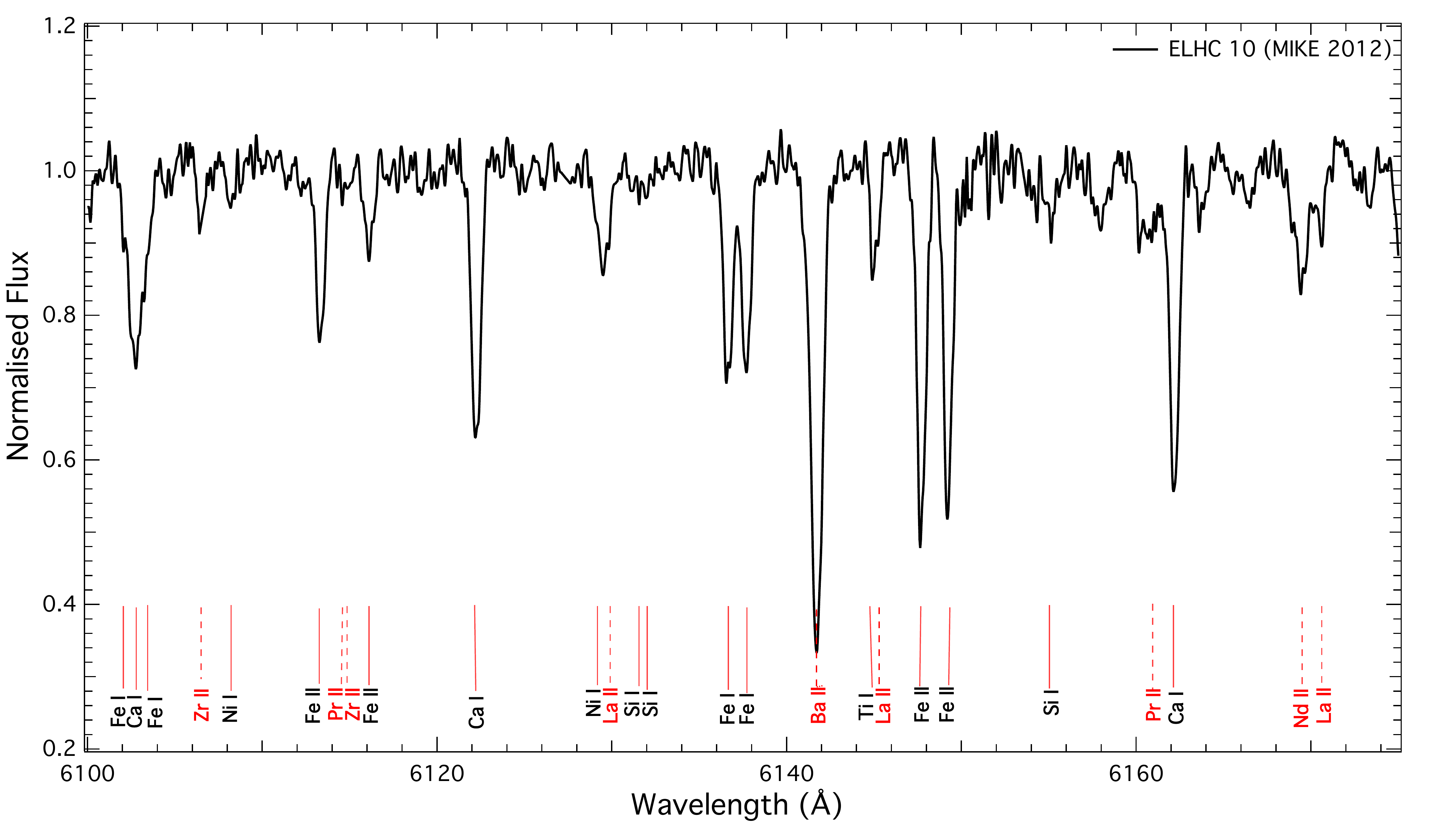}
    \caption{Complete line identification  of this spectral region using the VALD database. Elements with atomic number Z smaller that 30 are indicated with a solid line, light s-process elements are indicated with
    a dashed line. This spectrum at phase $\Phi_{o}$ = 0.47 was selected for abundance analysis due to the absence of DACs. It is shifted to match the laboratory wavelength of Ba\,{\sc ii} 6141.72 \AA. }
    \label{S-elements}
\end{figure*}

 \subsection{Signatures of s-process elements and abundance determination }\label{s-process}
 
The s-process nucleosynthesis is characterized by the presence of elements created by slow neutron capture reactions in the absorption features of elements such as  Ba\,{\sc ii} at  4554.03\,\AA,  5853.68\,\AA, 6141.72\,\AA, and 6496.89\,\AA, Sr {\sc ii} line at 4077.71\,\AA\ and 4215.52\,\AA\ (see, \citealt{Pereira2007} and \citealt{Kamath2014}, for further details on identifying signatures of s-process enhancements in post-AGB stars). 

The Ba\,{\sc ii}, Ba\,{\sc i} and Sr\,{\sc ii}  lines are present in our spectra and are stronger in the primary of ELHC\,10 than in the F-type supergiant UVES-POP template spectrum. This fact suggests that s-process enrichment has taken place in the primary. 
 In order to calculate the abundances of $\alpha$-elements and s-process species, we used the MIKE spectrum obtained on February 07, 2012 for analysis. This observation was carried out near maximum light at a phase ($\Phi=0.47$) corresponding to the superior conjunction of the post-AGB star, when its obscuration by the circumstellar disc and the contribution from secondary star should be minimal. Additional to this, the metallic lines at this phase do not present DACs and were nearly symmetric and hence easy to measure providing unblended lines (Fig. \ref{S-elements}).

The abundances are calculated using suitable programs coupled with
state-of-the-art atmospheric models \citep{Sneden1973, Kurucz1993} using the LTE approximation, as usual in the abundance analysis of post-AGB stars (e.g. \citealt{vanAarle2013}). 
We used as input parameters the effective temperature and surface gravity derived from our spectroscopic analysis (Section 3.2). Microturbulence velocity is obtained by removing any slope 
in the relation between the abundance from Fe\,{\sc i}  lines and the reduced EWs.
Abundances for all the other elements are obtained from EW measurements.
Usually cool stars show no evidence of significant rotation, so EWs are 
obtained from simple Gaussian fitting of spectral features. The NLTE effects should alter the ionization structure compared with the LTE model, but at the mild low metallicity and effective temperature of ELHC\,10 this effect should be less than 0.1 dex for Fe {\sc i} and 0.01 dex for Fe {\sc ii} \citep{Mashonkina2011}. In any case, the effective temperature derived by us comes from the comparison with the synthetic stellar spectra library, and does not depend on the LTE assumption.

 The final abundance analysis of the ELHC\,10 primary yields  the parameters shown in  Table\,7, where $Z$ is the proton number and $N$ represents the number of lines used for the abundance determination of the species. The iron abundance [Fe\,{\sc i} /\,H]$=-0.38$ is  consistent with a star located in the low metallicity environment of the LMC.
 
Derived abundances for the ELHC\,10 primary are plotted in Fig.   \ref{Abundance}. The simple mean of the [X/Fe] ratio for the $\alpha$-elements  Si, S, Ca and Ti, 
 is [$\alpha$/Fe]= $0.40\pm0.14$, normal when compared with Galactic objects in this metallicity range, and consistent with the  $\alpha$-enhancement  expected for LMC stars at their respective metallicities \citep{Pompeia2008, Van der Swaelmen2013}.  In Fig.  \ref{Abundance} we also compare the derived abundances of ELHC 10 with the abundances of J053253.51$-$695915.1, which is a post-AGB star confirmed in the LMC \citep{vanAarle2013}. It is clear that the primary of ELHC\,10 shows light s-process elements (magic neutron number 50) around Y and Zr which confirms the post-second dredge-up status of our star \citep{vanWinckel2003}.

As a by-product of our abundance analysis we calculated a microturbulence velocity of $\xi_{t}=6.8$ $\pm$ 0.5 km s$^{-1}$ and an upper limit for the projected rotational velocity of $20 \pm5$  km s$^{-1}$. 
The last figure comes from the width of the wings of the line profiles considering that a contribution of macro-turbulent velocity is also present.

\begin{table}\label{Abundance}
\caption{Abundances for ELHC\,10. $N$ represents the number  of lines used for the abundance determination of the species. The uncertainties in $\log \epsilon$, $\log \epsilon_{\odot}$, [X/H] and [X/Fe]
due to line to line scatter and model uncertainties were about 0.2 dex.}
\begin{tabular}{lcccccc}
 \hline
Species & $N$  & $\log \epsilon$      &  $\log \epsilon_{\odot}$ &  [X/H] &   [X/Fe] & $Z$\\\hline
C\,{\sc i}     & 2  & 8.72       & 8.52   &          0.20 &      0.58 & 6 \\
Si\,{\sc i}    & 1  & 7.56       & 7.55    &        0.01 &      0.39 &  14\\
S\,{\sc i}      & 2  & 7.36       &7.33      &       0.03  &     0.41 & 16\\
Ca\,{\sc i}     & 3  & 6.19       &6.36       &     -0.17  &     0.21 & 20 \\
Sc\,{\sc ii}    & 2  & 2.97      &3.17       &     -0.20 &      0.18 & 21 \\
Ti\,{\sc ii}     & 2  & 5.19       &5.02       &      0.17 &      0.55  & 22\\
Cr\,{\sc ii}     & 1  & 5.64       &5.67      &      -0.03 &      0.35&  24\\
Mn\,{\sc i}    & 2  & 5.05       &5.39       &     -0.34 &      0.04 & 25\\
Fe\,{\sc i}     & 15 & 7.12       &7.50      &      -0.38 &        \ldots& 26   \\
Zn\,{\sc i}     & 1  & 4.13       &4.60    &        -0.47 &     -0.09&  30\\
Y\,{\sc ii}   & 3 & 2.65  &    2.24      &      0.41  &    0.79 & 39 \\   
Zr\,{\sc ii}    & 1  & 2.73       &2.60    &         0.13 &      0.51 & 40\\
Ba\,{\sc ii}    & 1  & 2.79       &2.34    &         0.45 &      0.83 & 56\\
La\,{\sc ii}    & 1  & 1.67      &1.26  &         0.41 &      0.79 & 57\\
Nd\,{\sc ii}    & 1  & 2.30      &1.50     &        0.80 &      1.18  & 60\\
Eu\,{\sc ii}    & 1  & 1.03       &0.51     &        0.52 &      0.90 & 63\\\hline
\end{tabular}
\end{table}

\section{Discussion}

In this Section we calculate  the system mass function using results of our radial velocity analysis and discuss the possible scenarios for the system that are compatible with the existing data.

\begin{figure}
    \includegraphics[width=0.5\textwidth]{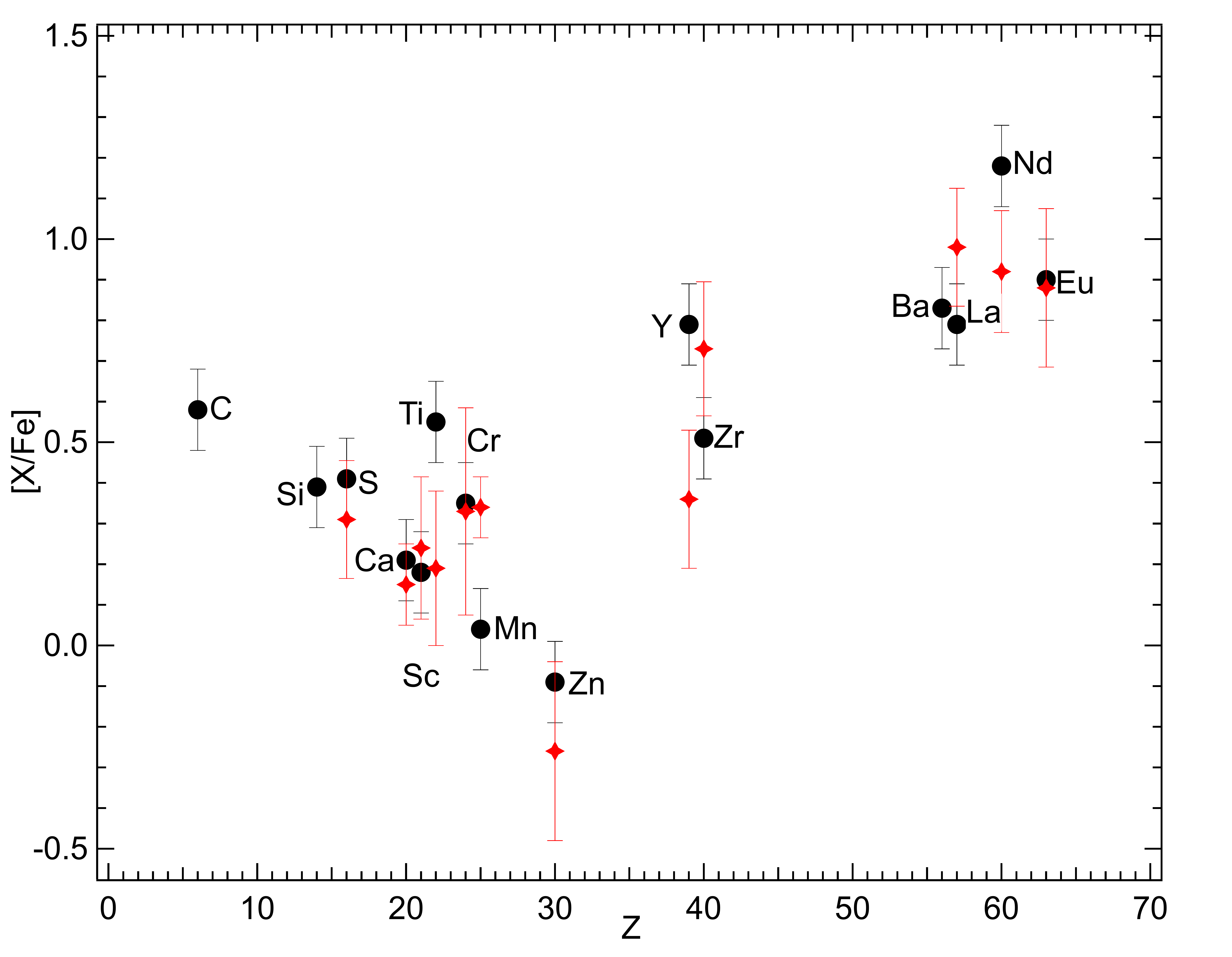}
    \caption{Derived abundance patterns for the primary of ELHC 10 (black points) in comparison with abundances for the same elements of the post-AGB star J053253.51$-$695915.1 (red points). Elements are labelled for clarity.}
    \label{Abundance}
\end{figure}

  \begin{figure}
    \includegraphics[width=0.5\textwidth]{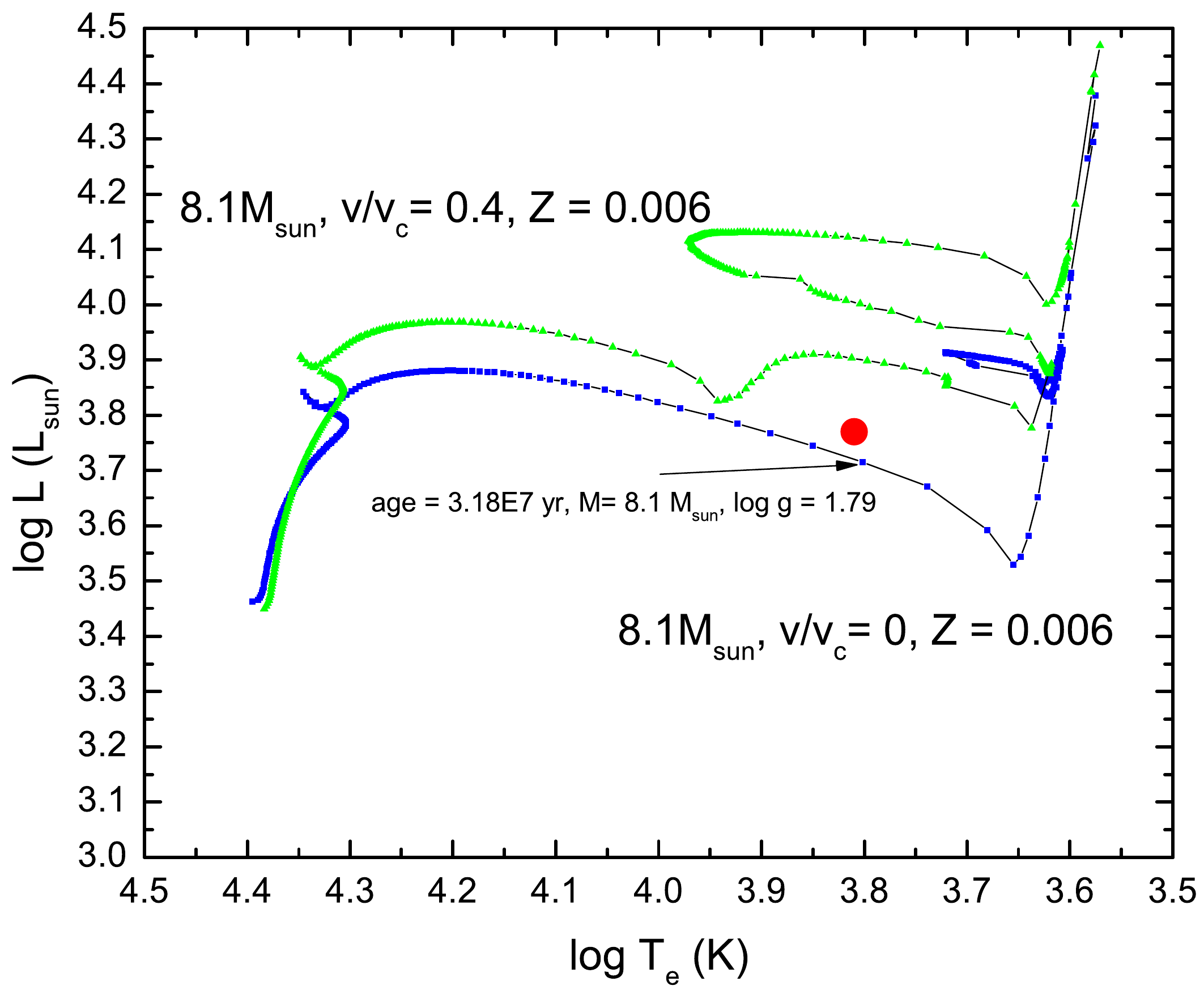}
    \caption{Position of ELHC~10 in the HR diagram. The location of ELHC~10 is shown with a red point in the diagram. The evolutionary tracks for single stars of initial mass 8.1 M$_{\odot}$ were taken from Georgy et al. (2013).
    We give the age, mass and surface gravity of the model closer to ELHC~10.}
    \label{HR_daiagram}
     \label{HRdiag}
\end{figure}

\begin{figure}
    \includegraphics[width=0.5\textwidth]{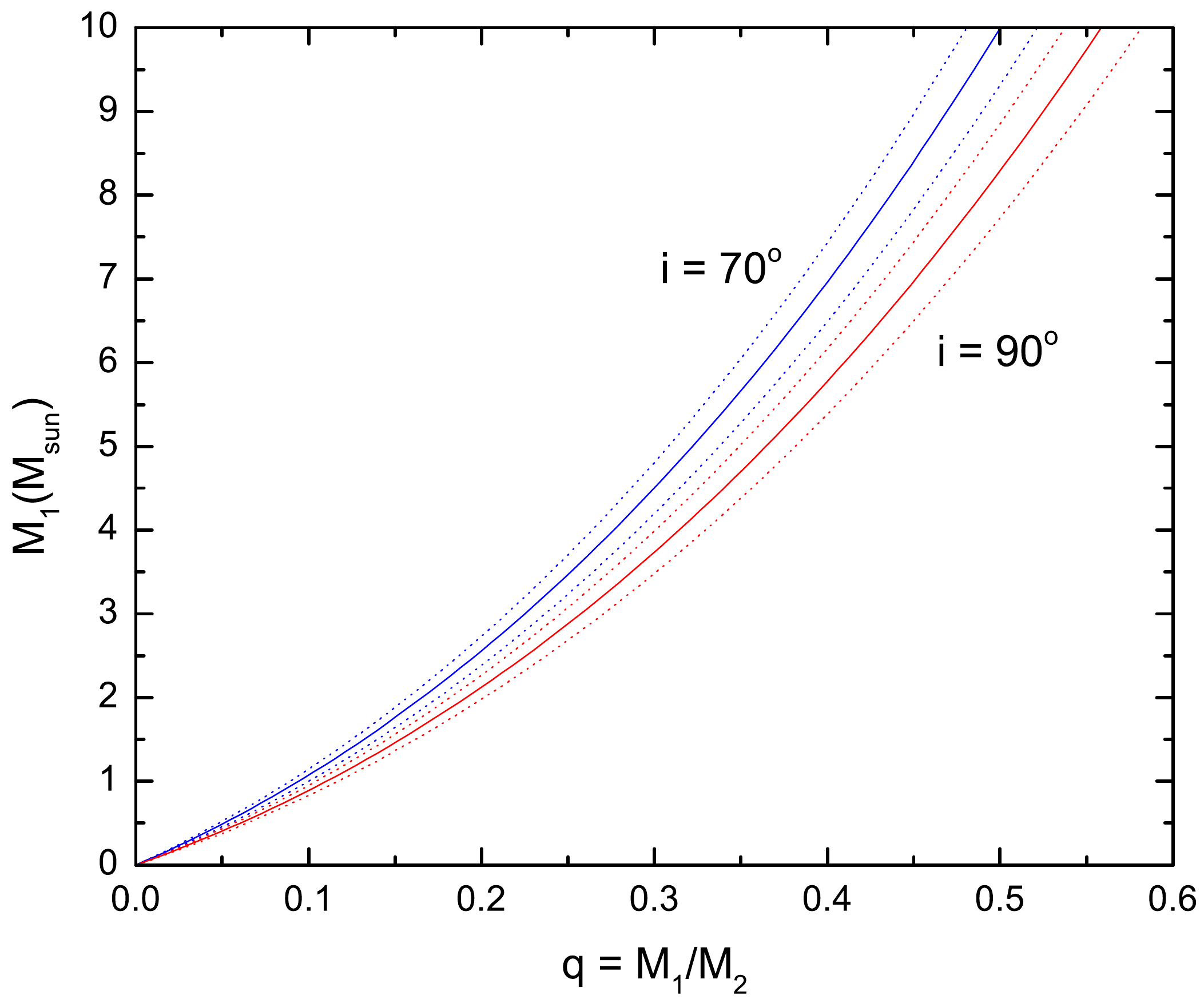}
    \caption{Primary mass versus mass ratio for different inclination angles according to the derived mass function. Dotted lines indicate the range due to the uncertainty in the mass function.}
    \label{massfunction}
\end{figure}

\subsection{On the nature of the primary star}

 Using the Stefan-Boltzmann law, the radius obtained from the SED fit and the spectroscopic temperature, the luminosity of the primary turns to be  $L=5 970\,L_{\odot}$. 
 Using $M_{v} = -4.7$ by de Wit et al. (2002) and neglecting the bolometric correction for a mid-F supergiant \citep{2000asqu.book.....C}  we get  $L=5 970\,L_{\odot}$ and the previous value is confirmed.
 We have compared the luminosity and temperature of the primary with the evolutionary tracks of single stars at metallicity $Z$ = 0.006 provided by Georgy et al. (2013). 
The primary closely fits the model of an initially (and present) 8.1 M$_{\odot}$ star at an age of $3.18\times 10^{7}$ yr but with a relatively large surface gravity log\,g = 1.79 (Fig.  \ref{HRdiag}).  This high mass primary 
 does not fit our results for the mass range and surface gravity inferred from the SED and the spectrum as given in Section 3.3. Moreover, it produces a problematic very high mass secondary 
 as we will show in the next Section.

\subsection{Constraints on the mass of the unseen secondary star and the semidetached case }\label{mass-ratio}

We calculate the mass function $f(m)$ defined for a single-lined spectroscopic binary as \citep{Hilditch2001}:


\begin{equation}\label{9}
f (m) = \frac{M_{1}sin^{3}i}{q(1+q)^{2}} = 1.0361\times10^{-7}\, K_{1}^{3}\, P\, \rm{M}_{\sun},
\end{equation}
  
\noindent 
where $P$ is the orbital period in days, $i$ is the inclination of the orbit and $K_{1}$ the primary radial velocity half-amplitude in km s$^{-1}$. 

Assuming an eccentricity $e= 0$ and using our derived $K_{1}$ we get $f(m)= 7.37$  $\pm$ 0.55 M$_{\odot}$. The possible masses for the secondary for a range of possible inclinations and primary masses, are shown in Fig.  \ref{massfunction}. 
It is clear that the secondary is always more massive that the primary.  Since it is not detected either in the continuum or  in the line spectrum probably it is hidden by the eclipsing structure.

We notice that, according to equation (6),  a primary of 8.1 M$_{\odot}$ implies a (unseen) secondary of at least 16 M$_{\odot}$. It is hard to explain why this high mass star does not heat the dark cool eclipsing structure. Let's remember  that to explain the equal depth in the eclipses at different bandpasses {\it and} the similar strength of  the lines inside/outside the main eclipse a dark
cool structure is required.

\begin{figure}
    \includegraphics[width=0.5\textwidth]{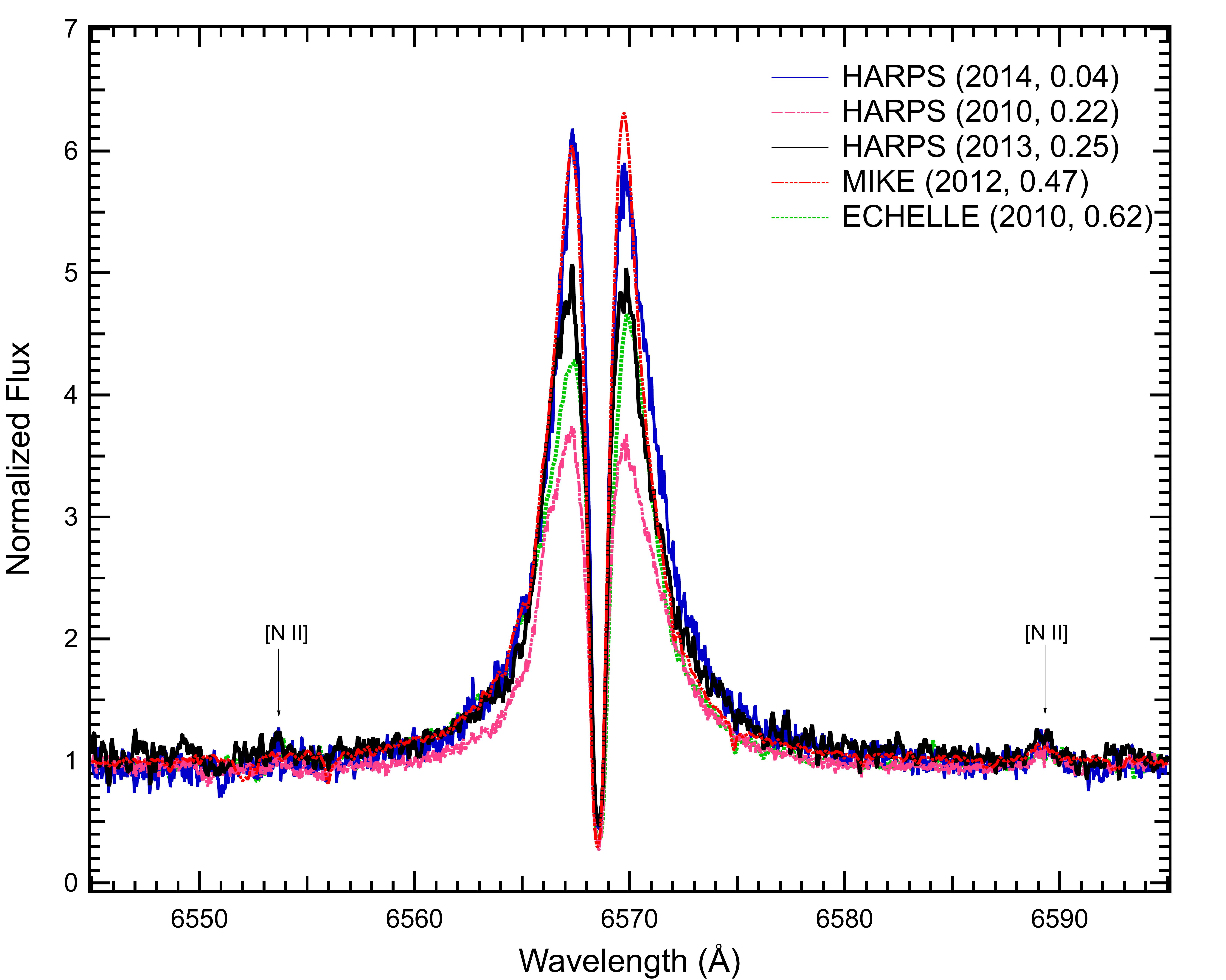}
    \caption{The HARPS (2010, 2013, 2014), MIKE (2012) and ECHELLE (2010) spectra showing  H$\alpha$ profile and N {\sc ii} emission lines. Fluxes are normalized to the continuum and heliocentric
    corrections have been applied.}
    \label{balmer}
\end{figure}

 Let's explore now the consequences of a possible semidetached configuration that should explain the existence of the eclipsing opaque structure as a disc produced by mass transfer due to Roche-lobe overflow.
Assuming  that the primary fills its Roche lobe, their mean density should be constrained by the orbital period for $q\leq 0.8$  as (in our nomenclature $q = M_{1}/M_{2}$):
 
 \begin{equation}\label{4}
\bar{\rho}\approx 110\,P_{0}^{-2} \mathrm{g\,cm^{-3}},
  \end{equation}
 where $P_{0}$ is the orbital period in hours (e.g. \citealt{Frank2002}).  For  ELHC\,10 we find $\bar{\rho}=3.9\times10^{-6}\ \mathrm{g\,cm^{-3}}$. Using this value, the radius derived in the previous section and assuming spherical symmetry  we calculate a mass for the primary $M_{1}\approx 0.625\ \mathrm{M_{\odot}}$. Surprisingly, this is very near the mass of a Post-AGB star with the observed luminosity.
In fact, using the  luminosity-core mass relation given by \citet{Wood1981} we obtain  $M_{1}=0.63\,M_{\odot}$,  
which fits the primary mass  restrictions derived in Section 3.3.
With this mass we get a mass-ratio $q= 0.07$ (for $i$ = 90$^{o}$) and  a mass for the unseen secondary star $M_{2}= 8.6$ \,M$_{\odot}$, consisting of a main-sequence B2\,V star.
We conclude that an hypothetical  semidetached configuration should imply a low-mass primary consistent with a Post-AGB classification,  as is discussed in the next section.

\subsection{Exploring the association with N120}

The association with N120 yields interesting clues on the history of the system.
N120 is a young association: the presence of H\,{\sc ii} regions and a SNR suggest an age of 
about 10\,Myr \citep{Laval1992}. This age imposes a problem for  the evolutionary stage of the low-mass primary. There has not been  enough time for the
low-mass primary to evolve to the observed supergiant stage. The problem disappears  if 
the primary has evolved from an initially  higher mass star in an evolutionary channel strongly influenced by the presence of the binary companion.
For instance,  the primary could have lost significant amounts of matter by Roche lobe overflow during  a semidetached stage as happens in Algols.  In this case, a primary of several solar masses
is expected in ELHC\,10 before mass ratio reversal, since the total mass is about 7 or 9 solar masses. On the contrary, a low mass progenitor is expected for the evolution of a single post-AGB star. 
Evolutionary tracks for single AGB stars have been published by \citet{Vassiliadis1994}. 
Extrapolating these models (that are for $Z=0.001$) to the mass of ELHC\,10 we find  mass for the progenitor of 1.5 $\mathrm{M_{\odot}}$ \citep{DeSmedt2012}. 
Possibly the evolution have followed a different path in the post-AGB star of ELHC\,10, because of the presence of the companion and epochs of significant mass transfer.  
This conjecture is consistent with the existence of massive post-AGB stars ($M \la 8 \mathrm{M_{\odot}}$; \citealt{Ventura2015}). 
 Since the amount of mass lost by the system during this process is hard to know, we cannot reconstruct the history 
of the binary and the evolutionary track up to the progenitors. 

We notice that the expected progenitor for a 0.62 M$_{\odot}$ post-AGB stars is  1.5 $\mathrm{M_{\odot}}$. 
It is impossible that this star has evolved to the current evolutionary stage {\it as a single star} if we assume  the same age as the nebular complex N120.




\subsection{H$\alpha$ emission and discs}\label{halpha}

H$\alpha$  is the most intense line in the  spectra and shows a double-peak emission, broad wings extending from 6557 to 6582 \AA\ and a mean separation of the blue and red peak of $\approx 115$   
km\,s$^{-1}$  with the blue peak approximately equal to the red peak.  The H$\alpha$ emission is characterized by several parameters whose values are given in Table\,8. The strong H$\alpha$ double-peak emission is consistent with 
the presence of an emitting disc. The central absorption component in the H$\alpha$ profile remains with the same radial  velocity in spite of changes in the radial velocity  of the primary over
the years (see Fig.\,\ref{balmer} and Table\, 8). The same occurs for the position of the violet and red emission peaks. The velocity of the central core  is  $-5\pm 1$ km\,s$^{-1}$, blue shifted from the systemic velocity, 
i.e. at velocity 260  km\,s$^{-1}$. 
The constancy of this velocity  is possible if (i) the disc is around a star considerably more massive than the primary, therefore it performs small (undetected) motions around the system center of mass or 
(ii) most of the emission comes from a slowly expanding circumbinary disc extending beyond the binary to considerable distances, producing a column density large enough to produce the almost stationary H$\alpha$ 
central absorption. In the following we give arguments favoring this last possibility.

Emission from a circumbinary disc is consistent with the fact  that the velocity of the H$\alpha$ central absorption  is practically the same as the velocity of
 the nebular complex N 120 C4 and C5 that are close to our star; viz. 260 km\,s$^{-1}$ (Section 2.2).
It is also consistent with the detection of stationary N\,{\sc ii} 6548, 6583 \AA\ forbidden emission lines, which are commonly attributed to low density material in the circumstellar environment. 
It is possible that the H$\alpha$ emitting disc extends into the interstellar medium to zones of low density where forbidden emission lines are produced. 
The circumbinary disc should be optically thin, to explain the steep Balmer emission decrement, and contribute with free-free emission to the system at infrared wavelengths (Fig.\,5). In addition, 
the large H$\alpha$ central absorption indicates strong self-absorption, this is possible if the disc is coplanar (or almost coplanar) with the binary. Finally, the disc is not eclipsed by the primary, as reveals the
H$\alpha$ equivalent width during secondary eclipse (Table 8).

\begin{table*}\label{Table3}
\caption{H$\alpha$ equivalent widths, intensities of the blue and red emission peaks normalized to the underlying continuum, intensity of the central depression,
RV of the overall profile, peak separation, full width at half-maximum and circumstellar reddening caused by  the H$\alpha$ emitting envelope  \citep{Dachs1988}.}
\begin{tabular}{llccccccccc}
\hline
Phase & Instrument & -EW & I$_{\mathrm{V}}$/I$_{\mathrm{Cont.}}$ & I$_{\mathrm{R}}$/\,I$_{\mathrm{Cont.}}$ & I$_{\mathrm{C}}$/\,I$_{\mathrm{Cont.}}$ &I$_{\mathrm{V}}$/\,I$_{\mathrm{R}}$ & RV$_{\mathrm{mean}}$ & $\Delta$ V$_{\mathrm{peak}}$ & FWHM & E$_{\mathrm{(B-V)}}^{\mathrm{CE}}$\\
 &   & (\AA)& & & & & (km\,s$^{-1}$) & (km\,s$^{-1}$) & (km\,s$^{-1}$) & (mag)\\\hline
 0.04 & HARPS (2014) & 25.5 & 6.19   & 5.91 &  0.42 &1.05 & 260 &  112 & 304  & 0.05\\
 0.22 & HARPS (2010) & 26.0 & 4.31 & 4.24  & 0.33 &1.02 & 260 & 118 & 269 & 0.05 \\
0.25 & HARPS (2013) & 22.8 & 5.07 & 5.04  & 0.45 &1.01 & 259 & 114 &  244&  0.05\\
0.47 & MIKE (2012) & 24.5  & 6.03   & 6.30 &0.29 &0.96 & 260 &112 & 247& 0.05\\
0.62 &  ECHELLE (2010) & 21.3 & 3.98 & 4.34 & 0.19 &0.92 & 262 &  110 & 234   &  0.04\\
\hline
\end{tabular}
\end{table*}

The circumbinary disc analyzed in this section is different from the opaque structure around the secondary.
In fact, Balmer emission can be produced in a relatively hot disc (T $\sim$ 10.000 $K$) but the opaque structure (probably another disc) around the secondary is cool; it does not contribute significantly 
to the flux in the optical region as evident from the constancy of line depths during the orbital cycle, even during eclipse  (Fig.\, \ref{FeBa_II}). 
 Coexistent circumstellar and circumbinary discs have been also found in some post-AGB binaries \citep{Gorlova2012, Gorlova2015, Hillen2013, Hillen2014, Hillen2015}.


\subsection{On the line splitting observed in ELHC\,10 }

Line splitting is a very particular property of ELHC\,10. It is observed in metallic lines, especially lines of Fe\,{\sc i},  Fe\,{\sc ii}  and Ti\,{\sc ii},
but disappears near secondary minimum at $\Phi_{o}$ = 0.47. Since these lines are stronger at this phase, it is possible that DACs and photospheric lines are blended. Line splitting are not observed in Ba\,{\sc ii}, Ca\,{\sc i}, Mg\,{\sc ii} and Si\,{\sc ii} lines except during main eclipse at phase 
 $\Phi_{o}$ = 0.04 (at least Ba\,{\sc ii} and Si\,{\sc ii} that are spectrally covered at this phase). Line splitting occurs in the form of DACs displaced to the red or blue side of the photospheric line.  
 

 Line splitting has been observed in lines of 
heavy elements in post-AGB stars with C-rich circumstellar environments and explained by
structured atmospheres \citep{2015AstL...41...14K}. For instance, in V\,5112\,Sgr 
the Ba\,{\sc ii} line profiles are split in multiple components \citep{Klochkova2013}. The case of ELHC\,10
is different, since its primary is not carbon-rich and the alternated visibility of 
DACs during the orbital cycle exclude any interpretation in terms of an structured envelope.  In fact, lines formed in an structured envelope around the primary should follow the primary motion during
the orbital cycle with a blue velocity shift, but not move from the red to the blue side of a line formed in the photosphere. In addition, we do not  observe DACs in ELHC~10 only in s-process elements, as
in V\,5112\,Sgr, but also in metallic lines of light elements. 

In ELHC\,10 DACs are probably formed by occultation of the primary photosphere by dense and extended gas regions co-rotating with the binary. This view is supported by the large strength of the features and their 
radial velocity behavior.  Since the number of available spectra is limited,  we cannot go deeper in this interpretation.

Evidence for interaction has been observed in the Galactic post-AGB star BD$+$46$^{o}$ 442,  an evolved binary with gas streams, jets and disc \citep{Gorlova2012}. 
Contrary to ELHC\,10, this last system shows strong orbital variability at the H$\alpha$ emission; this line has been interpreted
as formed in the disc around the secondary and the H$\alpha$ blue-shifted absorption in terms of a disc jet \citep{Gorlova2012}. \\

\subsection{ELHC\,10 in the context of post-AGB binaries}

70 high probability and 1337 candidate post-AGB LMC stars have been catalogued by \cite{vanAarle2011}. About half of them are probably binaries, a distinction made based on the shape of the spectral energy distribution,
indicating a stable and likely Keplerian disc. This is a surprising result, since one could expect that the residual material  from the AGB mass loss should be found in an expanding envelope. 
Therefore, it has been suggested that the formation of a circumbinary disc plays a major role in the evolution of post-AGB  systems \citep{2007BaltA..16..112V}.
Very few of these binaries have been studied in detail. To our knowledge, ELHC 10 could be the first post-AGB binary in the Large Magellanic Cloud, and the post-AGB binary with the highest progenitor mass and the highest-mass companion known.

Hitherto the process of disc formation in post-AGB stars remains obscure. For instance, 
\citet{Livio1988} and \citet{1998ApJ...500..909S}  showed that the disc can be the result of the AGB evolution after engulfing the companion with a common envelope. 
Other scenario proposes a disc-like structure formed by wind accretion \citep{Mastrodemos1999}. Finally, 
the disc can be the result of interaction between the AGB wind and the disc jet \citep{Akashi2008}.  While these scenarios can be possible in some systems at some evolutionary stages, the current observations of ELHC\,10 suggest that equatorial
mass loss through the outer Lagrangian points could form a circumbinary disc.  This is consistent with Roche-lobe overflow and the presence of a large and thick disc surrounding the secondary star.
The kind of outflows that possibly explain the DACs have been predicted  by hydrodynamical simulations of gas dynamics in close interacting binaries \citep{Bisikalo2003, Sytov2007}.  To accept this view,
we need to assume a non-standard evolution for the post-AGB star, as explained in Section 4.3.

\section[]{Conclusion}

In this paper we have studied the very interesting and complex eclipsing binary ELHC\,10, located in the young nebular complex N120 in the Large Magellanic Cloud.
While we have determined solid constrains for some parameters of the system, some others remains unbounded. The following is a list of our main conclusions:

\begin{itemize}

\item We find that ELHC\,10 is a SB1 long-period eclipsing binary with the main eclipse  produced by a dark structure hiding the secondary star. 
\item For the primary we determined an effective temperature of 6500 $\pm$ 250 $K$, log\,g  = 1.0 $\pm$ 0.5 and luminosity 5970 L$_{\sun}$.
\item From their radial velocities  we find a mass function of  7.37 $\pm$ 0.55 M$_{\sun}$.
\item If the primary fills its Roche lobe it also fits the core-luminosity relationship of a post-AGB star with a mass of  0.625 M$_{\sun}$. The abundance pattern is consistent with this classification.
\item A 8.1 M$_{\sun}$ primary fits the observed luminosity and temperature in a detached configuration, but implies a secondary of at least 16 M$_{\sun}$ and it is hard to explain why this high-mass star is not visible and does not heat their surrounding dark structure. Another problem with the high-mass primary is its late evolutionary stage, since the unseen secondary should be far more evolved, larger and luminous.
\item The very likely  membership to the young nebular complex N120 suggests that the only way for the primary to has arrived to the observed
advanced evolutionary stage (low-mass primary case) is from an initially much more massive star, something that could has occurred
by mass transfer due to Roche lobe overflow in a semidetached binary.
\item We find a prominent stationary H$\alpha$ double emission line which we attribute to a circumbinary disc. 
\item We discovered line splitting of metallic lines characterized by discrete absorption components (DACs) observed alternatively at the blue and red side of the photospheric
line profiles during the orbital cycle. 
\item These DACs are similar in appearance to those observed in carbon-rich post-AGB stars \citep{2015AstL...41...14K}, 
but cannot be interpreted as evidence for structured atmosphere as done by these authors for carbon-rich post-AGB systems. 
\item One possible interpretation for the DACs  observed in ELHC\,10 are gas streams in an interacting  binary. 

\end{itemize}

\section{Acknowledgments}

We thank W. Gieren, G. Pietrzy{\'n}ski and Araucaria-project observers for their help with data acquisition.  We also thank Dr. Claus Tappert for useful discussions on this object  and Dr. Willem-Jan de Wit by providing us the EROS light curves and also thanks for being the referee whose comments helped to improve a first version of this manuscript. This publication makes use of VOSA, developed under the Spanish Virtual Observatory project supported from the Spanish MICINN through grant AyA2008-02156. This research has made use of the SIMBAD database,
operated at CDS, Strasbourg, France. We
acknowledge the use of data from the UVES Paranal Observatory Project ESO DDT Program ID 266.D-5655.
R.E.M. acknowledges support by VRID-Enlace 214.016.001-1.0, Fondecyt 1110347 and the BASAL Centro de Astrof{\'{i}}sica y Tecnolog{\'{i}}as Afines (CATA) PFB--06/2007. This research was funded in part by the Ministry of Education, Science and Technological Development of Republic of Serbia through the project Stellar Physics (No. 176004). S.V. gratefully acknowledges the support provided by Fondecyt reg. 1130721.


\label{lastpage}
\end{document}